\documentclass%
[12pt]%
{article}
\usepackage{ifpdf}
\usepackage{pgf}
\pgfdeclareimage[width=15cm]{1a}{fig1aa}
\pgfdeclareimage[width=15cm]{2a}{fig2aa}
\usepackage{caption}
\usepackage{subfig}
\usepackage{subfloat}
\usepackage{amsmath}
\usepackage{amstext}
\usepackage{amsthm}
\usepackage{amssymb}
\usepackage{amsopn}
\usepackage{mathrsfs} 
\usepackage{dsfont} 
\usepackage[bottom]{footmisc}
\usepackage{indentfirst}
\usepackage{upref}
\usepackage{cite}
\usepackage{units}
\usepackage{fancybox}
\usepackage[lflt]{floatflt}
\numberwithin{equation}{section}
\renewcommand{\baselinestretch}{1.4}
\DeclareMathOperator{\tr}{Tr}
\DeclareMathOperator{\ee}{e}
\newcommand{\chib}{\chi_{\text{B}}}
\newcommand{\etan}{\eta_{\text{N}}}

\begin{document}
\begin{titlepage}
\renewcommand{\thefootnote}{\fnsymbol{footnote}} 
\renewcommand{\baselinestretch}{1.0}
\small\normalsize
\begin{flushright}
MZ-TH/08-23
\end{flushright}
\vspace*{3em}
\vspace{0.1cm}

\begin{center}   

{\LARGE \textsc{The role of Background Independence\\[2mm]for Asymptotic Safety in \\[6mm]Quantum Einstein Gravity\footnote{Talk given by M.R.\ at the WE-Heraeus-Seminar ``Quantum Gravity: Challenges and Perspectives'', Bad Honnef, April 14-16, 2008.}}}

\vspace{1.4cm}
{\large M.~Reuter and H.~Weyer}\\

\vspace{0.7cm}
\noindent
\textit{Institute of Physics, University of Mainz\\
Staudingerweg 7, D--55099 Mainz, Germany}\\

\end{center}

\vspace*{0.6cm}
\begin{abstract}
We discuss various basic conceptual issues related to coarse graining flows in quantum gravity. In particular the requirement of background independence is shown to lead to renormalization group (RG) flows which are significantly different from their analogs on a rigid background spacetime. The importance of these findings for the asymptotic safety approach to Quantum Einstein Gravity (QEG) is demonstrated in a simplified setting where only the conformal factor is quantized. We identify background independence as a (the\,?) key prerequisite for the existence of a non-Gaussian RG fixed point and the renormalizability of QEG.
\end{abstract}
\end{titlepage}
%
%
%
%
%
%
\setcounter{footnote}{0}
\section{Introduction}\label{s1}
Finding a logically consistent and predictive quantum theory of gravity continues to be one of the most challenging open problems in theoretical physics. Even though the recent years have seen considerable progress in loop quantum gravity, string theory, and asymptotic safety, to mention just three approaches \cite{kiefer}, it seems that certain essential ingredients of a satisfactory microscopic theory are still missing or only poorly understood. In any of these approaches one typically encounters problems which are conceptually very difficult and deep, and at the same time highly complex from the calculational point of view.
On the conceptual side, the most severe problem is perhaps the issue of background independence \cite{A,R,T}. Already classically General Relativity is distinguished from all other physical theories in that it does not only tell us how physical processes take place in a given spacetime but also describes the dynamics of spacetime itself. Many problems one encounters when searching for a quantum theory of gravity can be traced back to this crucial property of General Relativity, namely that it dynamically generates the ``arena'' in which all physics is going to take place. In particular, the mediator of the gravitational interaction, the metric or closely related fields, defines the proper length or mass scale of all dimensionful physical quantities.

\noindent 
\textbf{(A) Asymptotic Safety}~In the following we investigate a particular aspect of background (in)dependence which is particularly important in the context of asymptotic safety \cite{wein,mr,percadou,oliver1,frank1,oliver2,oliver3,oliver4,souma,frank2,prop,oliverbook,perper1,codello,litimgrav,frankmach,oliverfrac,jan1,jan2,creh1,creh2,je1,max,livrev}.
In this approach gravity is described by a quantum field theory of the metric tensor which is renormalized at a non-Gaussian renormalization group (RG) fixed point. This quantum field theory is defined by a functional RG trajectory on a ```theory space'' consisting of well-behaved diffeomorphism invariant action functionals. This trajectory must be \textit{complete} in the sense that it has neither an infrared nor an ultraviolet cutoff. In the ultraviolet the absence of unphysical divergences is guaranteed by the requirement that the trajectory must hit a fixed point there.

\noindent
\textbf{(B) Coarse graining in gravity}~In order to implement this idea one has to pick a concrete RG framework. In principle many choices are possible here; they differ by the generating functionals they employ, in the way field configurations get ``integrated out'' along the RG flow, and, related to that, the interpretation of the corresponding RG scale, henceforth denoted $k$. In theories on flat spacetime there exist implementations of the Wilsonian RG, the effective average action \cite{avact,ym,avactrev,ymrev} for instance, which have the special property that the mass scale $k$ has a ``quasi--physical'' meaning in the following sense: The basic functional RG equation (FRGE) describes the $k$-dependence of a family of effective actions $\left\{ \Gamma_k,~0 \leq k < \infty \right\}$ each of which defines an effective field theory valid near the scale $k$.

Going over to quantum gravity it is not clear a priori how one could introduce an RG scale with a comparable physical meaning. 
The problem is that if $k$ is to have the status of a physical momentum it must be a \textit{proper} rather than merely a \textit{coordinate} momentum. However, proper momenta, distances, or other dimensionful quantities require a metric for their definition, and if the metric is dynamical it is not clear with respect to which metric $k$ should be ``proper''. 
Proceeding naively, the average action of gravity would be a functional $\Gamma_k [g_{\mu \nu}]$ which, besides $k$, depends on a single argument $g_{\mu \nu}$. More precisely, $\Gamma_k [\,\cdot\,]$, for every fixed value of $k$, is a map from the space of metrics into the reals. This implies that from the point of view of $\Gamma_k [\,\cdot\,]$ with $k$ fixed all metrics have an equal status so that $k$ cannot be ``proper'' with respect to any particular one of them. This is a direct consequence of background independence. It entails that the naive implementation of the average action idea, leading to a family of functionals $\left\{ \Gamma_k \right\}$ which depend only on one metric argument, cannot be labeled by an RG scale with the above ``quasi--physical'' interpretation.

\noindent
\textbf{(C) The gravitational average action}~The actual effective average action for gravity constructed in \cite{mr} achieves the desired ``quasi--physical'' status of $k$ by using the background field technique. The idea is to fix an arbitrary background metric $\overline{g}_{\mu \nu}$, quantize the (not necessarily small) metric fluctuations $h_{\mu \nu}$ nonperturbatively in this background, and finally adjust $\overline{g}_{\mu \nu}$ in such a way that the expectation value of the fluctuation vanishes: 
$\overline{h}_{\mu \nu} \equiv \langle h_{\mu \nu} \rangle =0$.
In this way the background gets fixed dynamically. The advantage of this procedure is that the quantization can take advantage of many nonperturbative tools developed for field theories on non-dynamical backgrounds. More imortantly it is background independent in the sense that no special $\overline{g}_{\mu \nu}$ plays any distinguished role. During the quantization of the $h_{\mu \nu}$-field the background metric is kept fixed but is never specified explicitly.

In this construction the RG scale $k$ is ``proper'' with respect to the background metric. Technically one organizes the path integral over $h_{\mu \nu}$ according to eigenmodes of the covariant Laplacian $D^2 (\overline{g}_{\mu \nu})$ built from $\overline{g}_{\mu \nu}$ and cuts off the integration at the infrared (IR) scale $k^2$. This is done by adding a mode suppression term $\Delta_k S$ to the bare action. Hence $k$ is a $\overline{g}_{\mu \nu}$-proper momentum related to the scale set by the ``last mode integrated out'' and can be given an approximate physical meaning therefore. (See \cite{jan1,jan2} for a detailed discussion of this point.) This property of the gravitational average action is the central prerequisite for the effective field theory interpretation and for the possibility of performing ``RG improvements'' on the basis of $\Gamma_k$ \cite{bh,erick1,cosmo1,cosmo2,entropy,esposito,h1,h2,h3,girelli,litim,mof}.

The price one has to pay for this advantage is that the average action is now a functional of two metrics:
$\Gamma_k [g_{\mu \nu}, \overline{g}_{\mu \nu}] \equiv
\Gamma_k [\,\overline{h}_{\mu \nu};\overline{g}_{\mu \nu}]$.
Here $g_{\mu \nu} \equiv \overline{g}_{\mu \nu} 
+ \overline{h}_{\mu \nu}$ is the expectation value of the microscopic metric. Only after having solved for the (now more complicated) RG flow of 
$\Gamma_k [g_{\mu \nu}, \overline{g}_{\mu \nu}]$ one can impose
$\overline{h}_{\mu \nu} =0$ and define the reduced functional
$\overline{\Gamma}_k [g_{\mu \nu}] \equiv
\Gamma_k [g_{\mu \nu}, g_{\mu \nu}]$
which generates the same on-shell matrix elements as the original one \cite{back}.

\noindent
\textbf{(D) ``Background independence'' via background field technique}~It should be stressed that the average action $\Gamma_k [\,\cdot\, , \, \cdot \,]$ and its RG flow are ``\textit{background independent}'' objects, in the sense of the word as it is used in loop quantum gravity \cite{A,R,T}, for instance. Both metrics, $g_{\mu \nu}$ and $\overline{g}_{\mu \nu}$, are just freely variable arguments and no metric plays any distinguished role\footnote{Here and in the following the term ``background independence'', put in quotation marks, means the absence of a preferred metric. Referring to the background field formalism, no quotation marks will be used.}. Furthermore, the mode cutoff is defined in terms of $D^2 (\,\overline{g}_{\mu \nu})$ which involves the variable metric $\overline{g}_{\mu \nu}$ and not any rigid one. This is in sharp contrast to matter field theories on a non-dynamical spacetime with a metric $g^{\text{non-dyn}}_{\mu \nu}$. There $\Delta_k S$ is constructed from $D^2 (g^{\text{non-dyn}}_{\mu \nu})$ which does indeed single out a specific metric. The resulting flow is not ``background independent'' in the above sense.

Besides fixing the physical scale of $k$, the use of the background field technique has a second, conceptually completely independent advantage: If one employs a gauge fixing term which is invariant under the background gauge transformations the resulting average action is a \textit{diffeomorphism invariant} functional of its arguments \cite{back}.

In the construction of the gravitational average action in \cite{mr} these two issues are intertwined and because of the complexity of realistic RG flows it is not easy to see how precisely the $\overline{g}_{\mu \nu}$-dependence of the IR cutoff
$\Delta_k S [h_{\mu \nu}; \overline{g}_{\mu \nu}]$ influences the flow. We shall describe this influence in a setting as ``clean'' as possible, namely in an approximation to the full gravitational RG flow where gauge issues play no role and the impact of this $\overline{g}$-dependence of the cutoff can be studied in isolation. The implications of the $\overline{g}$-dependence are at the very heart of quantum gravity. It arises only because the metric has the crucial property, not shared by any other field, of defining the proper size of all dimensionful quantities, including that of $k$.

Within a different theory of gravity, and in a different formal setting, Floreanini and Percacci \cite{floper} have made similar observations. They studied a perturbatively renormalizable gauge theory of vielbein and spin connection fields. While asymptotic safety is not an issue there, they demonstrated that the quantization of the model results in a ``bimetric theory'', and depending on which metric is used in the ultraviolet (UV) regulator different effective potentials are obtained for the conformal factor.

\noindent
\textbf{(E) The conformally reduced theory}~The system we are going to study in the following obtains by approximating the gravitational RG flow in two ways: First, we restrict the theory space to that of the familiar Einstein--Hilbert truncation whose RG flow is known in full generality \cite{mr,frank1}. Second, we quantize only the conformal factor of the metric but not the other degrees of freedom it carries. This ``conformally reduced Einstein--Hilbert'' (or ``CREH'') truncation leads to a modified RG flow on the same theory space as the full Einstein--Hilbert truncation, and it will be very instructive to compare the two.

All metrics appearing in the CREH framework, the integration variable in the path integral, $\gamma_{\mu \nu}$, as well as $\overline{g}_{\mu \nu}$ and $g_{\mu \nu}$, are of the type ``conformal factor times $\widehat{g}_{\mu \nu}$'' where $\widehat{g}_{\mu \nu}$ is a reference metric which is never changed; for example, $\widehat{g}_{\mu \nu} = \delta_{\mu \nu}$. In this way, $\gamma_{\mu \nu}$, $\overline{g}_{\mu \nu}$,
and $g_{\mu \nu}$ get represented by a single ``scalar'' function, their respective conformal factor. The background metric, for instance, is written as $\overline{g}_{\mu \nu} = 
\chib^2 (x) \, \widehat{g}_{\mu \nu}$. If one inserts the metric $\phi^2 \, \widehat{g}_{\mu \nu}$ into the Einstein--Hilbert action one obtains a $\phi^4$-type action for the field $\phi$, with a $\phi^4$-coupling proportional to the cosmological constant. We shall analyze this scalar--looking theory by means of an effective average action. We use a background approach which is analogous to the one used in the full gravitational FRGE. In particular the conformal factor of $\overline{g}_{\mu \nu}$ sets the physical scale of $k$. So, conceptually, this simplified setting is exactly the same as in the full gravitational flow equation, the only difference is that we allow only the quantum fluctuations of the conformal factor to contribute to the RG running of the couplings, i.\,e.\ the Newton and the cosmological constant, respectively.

\noindent
\textbf{(F) Conformal factor vs.\ $\boldsymbol{\phi^4}$-theory}~The standard quantization of $\phi^4$-theory by means of an FRGE for the average action is fairly well understood \cite{avactrev}. It amounts to using a $\overline{g}_{\mu \nu}$-independent cutoff. Here $\Delta_k S$ is built from $\widehat{g}_{\mu \nu}$ which is usually taken to be the metric of flat Euclidean space. It is this metric $\widehat{g}_{\mu \nu}$ which defines the meaning of $k$. This scheme is the natural one when $\phi$ is a conventional scalar matter field. By now a lot is known about the resulting RG flow \cite{avactrev}. In particular, above all mass thresholds one recovers the $\ln (k)$-running of the $\phi^4$-coupling which is familiar from perturbation theory.

If $\phi$ is the conformal factor of the metric the situation is different. Now it is natural to define $\Delta_k S$ and hence $k$ in terms of the \textit{adjustable} background metric $\overline{g}_{\mu \nu} = 
\chib^2 (x) \, \widehat{g}_{\mu \nu}$; its conformal factor $\chib$ is determined dynamically by the condition that the fluctuations about $\chib$ have vanishing expectation value. We shall see that the resulting RG flow is quite different from the standard scalar one. Typically one finds that the RG running is much faster in the gravitational case. 

For instance, there is a regime where the slow $\ln (k)$-running of the standard scalar is replaced by a much stronger $k^4$-running of the $\phi^4$-coupling. In this regime the $\phi^4$-coupling is proportional to the cosmological constant, $\Lambda_k$. Hence, in this particular regime, $\Lambda_k \propto k^4$.
This quartic cutoff dependence is something very well known, of course. It is precisely what one finds by summing zero-point energies, or rediscovers as quartic divergences in ordinary Feynman diagram calculations. Moreover it agrees with the result from the full Einstein--Hilbert truncation.

To summarize this important point: The (expected) behavior $\Lambda_k \propto k^4$ obtains only if we respect ``background independence'' and appreciate the very special role of gravity, namely that it determines all proper scales, including that of $k$. We find $\Lambda_k \propto k^4$ only if we define the cutoff with $\overline{g}_{\mu \nu} = 
\chib^2 (x) \, \widehat{g}_{\mu \nu}$, while we obtain the much weaker $k$-dependence $\Lambda_k \propto \ln (k)$ if we treat $\phi$ as an ordinary scalar.

Earlier on Polyakov \cite{polyakov} and Jackiw et al.\ \cite{jackiw} have pointed out that in the CREH approximation the gravitational action is of the $\phi^4$-type and argued on the basis of standard scalar field theory that the cosmological constant should have a logarithmic scale dependence therefore. Our results indicate that if one wants to attach a physical meaning to $k$ by measuring it in units of $\phi$ itself the running of $\Lambda_k$ is much faster in fact.

\noindent
\textbf{(G) Asymptotic safety in a $\boldsymbol{\phi^4}$-type theory}~Perhaps the most unexpected and striking feature of the CREH flow is that it admits a non-Gaussian RG fixed point (NGFP) with exactly the same qualitative properties as the one in the full Einstein--Hilbert truncation. The comparatively simple dynamics of a $\phi^4$-theory is enough to achieve asymptotic safety \textit{provided one quantizes the theory in a ``background independent'' way}.

At the NGFP the cosmological constant is positive and this translates to a negative $\phi^4$-coupling. Long ago Symanzik \cite{syman,hist} showed that the scalar $\phi^4$-theory with a negative coupling constant is asymptotically free; its coupling strength vanishes logarithmically at high momenta. Using the cutoff appropriate for the gravitational field the asymptotically free RG flow becomes an asymptotically safe one, a NGFP develops.

The investigations using the gravitational average action which have been performed during the past few years \cite{mr,percadou,oliver1,frank1,oliver2,oliver3,oliver4,souma,frank2,prop,oliverbook,perper1,codello,litimgrav} indicate that full Quantum Einstein Gravity (QEG) is indeed likely to possess a NGFP which makes the theory asymptotically safe. Increasingly complicated truncations of theory space were analyzed whereby all modes of the metric were retained. The results which we shall describe in the following indicate that the NGFP that was found in these analyses is perhaps easier to understand than it was thought up to now. It seems that, to some extent, it owes its existence to an essentially ``kinematical'' phenomenon which is related to the requirement of ``background independence'' and the fact that the dynamical field itself, the metric, determines the proper value of the coarse graining scale. The complicated selfinteractions of the helicity-2 modes, on the other hand, can be omitted without destroying the NGFP. While also characteristic of gravity, they seem not to be essential for asymptotic safety.

The remaining sections of this paper are organized as follows. As a preparation we discuss in Section \ref{s2} the conformally reduced Einstein--Hilbert action. Then, in Section \ref{s3}, we derive an exact flow equation for conformally reduced gravity. In Section \ref{s4} we specialize it for the CREH truncation and explain in particular the conceptual differences of the theory presented here and standard scalar matter field theories. In Section \ref{s5} we analyze the RG equations obtained from the CREH truncation and show that they predict a NGFP. The conclusions are contained in Section \ref{s6}.

For further details we refer the reader to \cite{creh1} and \cite{creh2}.
%
%
%
%
%
%
\section{The Conformally Reduced Einstein--Hilbert Action}\label{s2}
In $d$ spacetime dimensions, the Euclidean Einstein--Hilbert action reads
\begin{align} \label{2.1}
S_{\text{EH}} [g_{\mu \nu}]
& =
- \frac{1}{16 \pi \, G} \,
\int \!\!\mathrm{d}^d x~\sqrt{g\,} \,
\bigl( R (g) - 2 \, \Lambda \bigr).
\end{align}
Henceforth we shall assume that the argument $g_{\mu \nu}$ is a conformal factor times a fixed, non-dynamical reference metric $\widehat{g}_{\mu \nu}$. We would like to parameterize this conformal factor in terms of a ``scalar'' function $\phi (x)$ in such a way that the kinetic term for $\phi$ becomes standard,
$\propto (\partial_\mu \phi)^2$. This is indeed possible for any dimensionality. Introducing $\phi$ according to \cite{jackiw},
\begin{align} \label{2.2}
g_{\mu \nu} 
& =
\phi^{2 \nu (d)} \, \,\widehat{g}_{\mu \nu},
\end{align}
with the exponent
\begin{align} \label{2.3}
\nu (d) & \equiv \frac{2}{d-2}
\end{align}
standard formulas for Weyl rescalings yield the following result for $S_{\text{EH}}$ evaluated on metrics of the form \eqref{2.2}:
\begin{align} \label{2.4}
S_{\text{EH}} [\phi]
& =
- \frac{1}{8 \pi \, \xi (d) \, G} \,
\int \!\! \mathrm{d}^d x~ \sqrt{\widehat{g} \,} \,
\left(
\tfrac{1}{2} \, \widehat{g}\,^{\mu \nu} \, \partial_\mu \phi \, 
\partial_\nu \phi
+ \tfrac{1}{2} \, \xi (d) \, \widehat{R} \, \phi^2
- \xi (d) \, \Lambda \, \phi^{2d/(d-2)} 
\right).
\end{align}
Here $\widehat{R}$ is the curvature scalar of the reference metric $\widehat{g}_{\mu \nu}$, and
\begin{align} \label{2.5}
\xi (d) & \equiv \frac{d-2}{4 \, \left( d-1 \right)}.
\end{align}
In $4$ dimensions we have $\nu = 1$ and $\xi = 1/6$ so that the choice
\begin{align} \label{2.6}
g_{\mu \nu} & = \phi^2 \, \,\widehat{g}_{\mu \nu}
\end{align}
converts the Einstein--Hilbert action to a kind of ``$\phi^4$-theory'':
\begin{align} \label{2.7}
S_{\text{EH}} [\phi]
& =
- \frac{3}{4 \pi \, G} \,
\int \!\! \mathrm{d}^4 x~ \sqrt{\widehat{g} \,} \,
\left(
\tfrac{1}{2} \, \widehat{g}\,^{\mu \nu} \, \partial_\mu \phi \, 
\partial_\nu \phi
+ \tfrac{1}{12} \, \widehat{R} \, \phi^2
- \tfrac{1}{6} \, \Lambda \, \phi^4 
\right).
\end{align}
We shall refer to the action \eqref{2.4} and its special case \eqref{2.7} as the conformally reduced Einstein--Hilbert or ``CREH'' action.

Up to now $\widehat{g}_{\mu \nu}$ is an arbitrary metric, defined on the same smooth manifold as $g_{\mu \nu}$. Later on we shall fix the topology of this manifold to be that of flat space $\mathrm{R}^d$ or of the sphere $\mathrm{S}^d$.

For $d>2$, the case we shall always assume in the following, the kinetic term in $S_{\text{EH}} [\phi]$ of eq.\ \eqref{2.4} is always negative definite due to the ``wrong sign'' of its prefactor. As a result, the action is unbounded below: for a $\phi (x)$ which varies sufficiently rapidly $S_{\text{EH}} [\phi]$ can become arbitrarily negative. This is the notorious conformal factor instability.

Leaving aside issues related to the functional measure, quantizing gravity in the CREH approximation based upon the bare action $S_{\text{EH}} [\phi]$ is similar to quantizing a scalar theory with an action of the general type
\begin{align} \label{2.8}
S [\phi]
& =
c \, \int \!\! \mathrm{d}^4 x ~
\Big\{ - \tfrac{1}{2} \left( \partial \phi \right)^2
+ U (\phi) \Big\}
\end{align}
where $c$ is a positive constant.
For the sake of the argument let us assume that 
$\widehat{g}_{\mu \nu} = \delta_{\mu \nu}$ is the flat metric on $\mathrm{R}^4$. Then $S_{\text{EH}}$ of \eqref{2.7} is indeed of the form \eqref{2.8} with the potential 
$U (\phi) = \tfrac{1}{6} \, \Lambda \, \phi^4$ and
$c = 3/ (4 \pi G) >0$. Let us assume that the cosmological constant is positive, the case which will be relevant later on. For $\Lambda >0$ the potential term in the action \eqref{2.8} is positive definite, while the kinetic piece is negative definite. We would like to explore the quantum theory based upon the functional integral
\begin{align} \label{2.9}
I & \equiv \int \!\! \mathcal{D} \phi~
\ee^{i \widetilde{S} [\phi]}
\end{align}
where $\widetilde{S}$ is the Wick rotated version of $S$, with
$(\partial \phi)^2 \equiv \eta^{\mu \nu} \, \partial_\mu \phi \,
\partial_\nu \phi$. One would expect that in this theory the wrong sign of the kinetic term drives the condensation of spatially inhomogeneous ($x$-dependent) modes, i.\,e.\ the formation of a ``kinetic condensate'' similar to the one discussed in \cite{kincond}. The amplitude of the inhomogeneous modes cannot grow unboundedly since this would cost potential energy.

Next let us look at the closely related theory with the ``inverted'' action $S_{\text{inv}} [\phi] \equiv - S [\phi]$. Thus
\begin{align} \label{2.10}
S_{\text{inv}} [\phi]
& =
c \, \int \!\! \mathrm{d}^4 x ~
\Big\{ + \tfrac{1}{2} \left( \partial \phi \right)^2
+ V (\phi) \Big\}
\end{align}
with the negative potential
\begin{align} \label{2.11}
V (\phi) & \equiv - U (\phi) \leq 0.
\end{align}
In pulling out a global minus sign from $S$ the instability inherent in the theory has been shifted from the kinetic to the potential term. According to $S_{\text{inv}}$, the kinetic energy assumes its minimum for homogeneous configurations $\phi = const$, but the inverted potential $V (\phi) = - \tfrac{1}{6} \, \Lambda \, \phi^4$ becomes arbitrarily negative for large $\phi$.

Even though $S$ and $S_{\text{inv}}$ appear to be plagued by instabilities of a very different nature, they nevertheless describe the same physics (up to a time reflection). The path integrals involving $S$ and $S_{\text{inv}}$ are related by a simple complex conjugation:
\begin{align} \label{2.12}
I_{\text{inv}}
& \equiv 
\int \!\! \mathcal{D} \phi~
\ee^{- i \widetilde{S} [\phi]}
= I^\ast.
\end{align}
We shall refer to the formulation in terms of $S$ and $S_{\text{inv}}$ as the \textit{original picture} and the \textit{inverted picture}, respectively.

So we see that for \textit{pure} gravity in the CREH approximation the ``wrong'' sign of the kinetic term can be traded for an upside down potential. The FRGE formalism we are going to develop will effectively correspond to the inverted picture. As we shall see it is indeed the $\Lambda>0$ case that will be relevant to asymptotic safety. Hence the conformal factor dynamics is described by an action with positive kinetic but negative potential term.

Interestingly enough, this kind of $\phi^4$-theory with a negative coupling constant was discussed by Symanzik \cite{syman} long ago. He showed that the coupling strength vanishes at short distances, thus providing the first example of an asymptotically free quantum field theory \cite{hist}.
%
%
%
\section{Effective Average Action for the Conformal Factor}\label{s3}
\subsection{The Background Field Method}\label{s3.1}
The most important difference between the conformal factor and an ordinary scalar is that $\phi$ determines the magnitude of all physical scales; in particular it determines the proper scale that is to be ascribed to a given numerical value of the IR cutoff $k$ appearing in the FRGE context. For this reason the quantization of $\phi$ by means of an FRGE differs from the standard one. In fact, even though gauge issues do not play any role here, the background field method has to be employed. This approach will allow us to give a meaning to statements like ``$\Gamma_k$ describes the dynamics of fields averaged over spacetime volumes of extension $\sim k^{-1}$''
in presence of a quantized metric where a priori it is unclear in which metric the extension of those spacetime volumes is measured.

Before we can set up the RG formalism we must explain the background--reformulation of the path integral underlying the quantum field theory of the conformal factor. We start from a formal path integral\footnote{Since this is customary in the literature we shall use a Euclidean notation in the general discussions. At the formal level it is trivial to obtain the corresponding Lorentzian formulas by replacing $-S \to iS$, etc.; for the time being the positivity properties of $S$ play no role.}
\begin{align} \label{3.1}
\int \!\! \mathcal{D} \chi ~\ee^{-S [\chi]}
\end{align} 
where $S$ is an arbitrary bare action (perhaps related, but not necessarily identical to $S_{\text{EH}}$) and $\chi (x)$ denotes the microscopic (``quantum'') conformal factor field. (The notation $\phi (x)$ will be reserved for its expectation value.) We think of \eqref{3.1} as descending from a path integral over quantum metrics $\gamma_{\mu \nu} (x)$,
\begin{align} \label{3.2}
\int \!\! \mathcal{D} \gamma_{\mu \nu}~
\ee^{-S_{\text{grav}} [\gamma_{\mu \nu}]},
\end{align}
by a restriction to metrics of the form
\begin{align} \label{3.3}
\gamma_{\mu \nu} & = \chi^{2 \nu} \, \,\widehat{g}_{\mu \nu}.
\end{align}
The integrals \eqref{3.1} and \eqref{3.2} refer to a spacetime manifold of a given topology and $\widehat{g}_{\mu \nu}$ is a reference metric consistent with this topology. The action $S [\chi]$ depends parametrically on $\widehat{g}_{\mu \nu}$ but we shall not indicate this dependence notationally. The non-dynamical, classical metric $\widehat{g}_{\mu \nu}$ is considered fixed once and for all; it has no analog in the full theory and is not to be confused with the background metric and the corresponding conformal factor which we introduce next.

We decompose the variable of integration, $\chi$, as the sum of a classical, fixed background field $\chib$ and a fluctuation $f$:
\begin{align} \label{3.4}
\chi (x) & = \chib (x) + f (x).
\end{align}
Even though we frequently use the term ``fluctuation'', $f (x)$ is not assumed small, and no expansion in powers of $f (x)$ is performed here.
We assume that the measure $\mathcal{D} \chi$ is translational invariant so that \eqref{3.1} can be replaced by $\int \! \mathcal{D} f \, \exp (-S[\chib +f])$. Actually it is sufficient to assume that the original $\mathcal{D} \chi$ equals a translational invariant measure up to a Jacobian since we may include the logarithm of this Jacobian in $S$.

At this point it is natural to introduce a background--type generating functional by coupling an external source $J (x)$ to the fluctuation only:
\begin{align} \label{3.5}
\exp \bigl( W [J; \chib] \bigr)
& =
\int \!\! \mathcal{D} f ~
\exp \left( -S [\chib +f]
+ \int \!\! \mathrm{d}^d x~\sqrt{\widehat{g}\,} \,
J (x) \, f (x) \right).
\end{align}
Repeated differentiation of $W$ with respect to the source yields the connected $n$-point functions of $f$ in presence of $J$. In particular the normalized expectation value of the fluctuation is
\begin{align} \label{3.6}
\overline{f} (x) & \equiv \langle f (x) \rangle
= \frac{1}{\sqrt{\widehat{g} (x)\,}\,} \,
\frac{\delta W[J; \chib]}{\delta J (x)}.
\end{align}
The field thus obtained is functionally dependent on both $J$ and $\chib$, i.\,e.\ $\overline{f} = \overline{f} [J; \chib]$. We assume that this relationship can be solved for the source, $J = J[\,\overline{f}; \chib]$, and introduce the Legendre transform of $W$:
\begin{align} \label{3.7}
\Gamma [\,\overline{f}; \chib]
& =
\int \!\! \mathrm{d}^d x~\sqrt{\widehat{g}\,} \,
J [\,\overline{f}; \chib] (x) \, \overline{f} (x)
- W \bigl[ J [\,\overline{f}; \chib]; \chib \bigr].
\end{align}
This definition implies the effective field equation
\begin{align} \label{3.8}
\frac{\delta \Gamma [\,\overline{f}; \chib]}{\delta \overline{f} (x)}
& = J (x).
\end{align}
More generally, repeated differentiation of $\Gamma$ with respect to $\overline{f} (x)$ yields the 1PI $n$-point correlators of $\overline{f}$ in presence of $J$. The source can be ``switched off'' by equating $\overline{f}$ after the differentiations to the function $\overline{f}_0 [\chib] (x) \equiv \overline{f} \,[J=0; \chib] (x)$. Note that $\overline{f}_0$ has no reason to vanish in general, and that the resulting $n$-point functions still depend on $\chib$. The expectation value of the complete conformal factor reads
\begin{align} \label{3.9}
\phi & \equiv \big\langle \left( \chib + f \right) \big\rangle
= \chib + \overline{f},
\end{align}
and sometimes it will be convenient to regard $\Gamma$ a 
functional of $\phi$ and $\chib$ rather than $\overline{f}$ and $\chib$:
\begin{align} \label{3.10}
\Gamma [\phi, \chib] & \equiv 
\Gamma [\,\overline{f} = \phi - \chib;\,\chib].
\end{align}
For the restriction of this function to equal arguments $\phi = \chib$ which amounts to a vanishing fluctuation expectation value we write
\begin{align} \label{3.11}
\overline{\Gamma} \,[\phi] & \equiv \Gamma [\phi, \phi]
= \Gamma [\,\overline{f} =0;\,\chib=\phi].
\end{align}

It is instructive to compare the above generating functionals in the background approach with those in the standard (``st''), i.\,e.\ non-background formalism. There one would define $W_{\text{st}} [J]$ by
\begin{align} \label{3.12}
\exp \bigl( W_{\text{st}} [J] \bigr)
& =
\int \!\! \mathcal{D} \chi \,
\exp \left( - S [\chi] 
+ \int \!\! \mathrm{d}^d x~ \sqrt{\widehat{g}\,} \,
J (x) \, \chi (x) \right)
\end{align}
and the standard effective action $\Gamma_{\text{st}} [\phi]$ would obtain as the Legendre transform of $W_{\text{st}} [J]$. Exploiting the translational invariance of $\mathcal{D} \chi$ it is easy to see that the two sets of functionals are related in a rather trivial way:
\begin{gather}
\label{3.13}
W [J; \chib] 
= 
W_{\text{st}} [J] - \int \!\! \mathrm{d}^d x~\sqrt{\widehat{g}\,} \,
J (x) \, \chib (x)
\\
\label{3.14}
\Gamma [\,\overline{f}; \chib]
=
\Gamma_{\text{st}} [\chib+\overline{f}\,]
\quad \Longleftrightarrow \quad
\Gamma [\phi, \chib] = \Gamma_{\text{st}} [\phi]
\\
\label{3.15}
\overline{\Gamma}\, [\phi] 
= \Gamma_{\text{st}} [\phi].
\end{gather}
The key property of the background formalism is that the standard $n$-point functions
\begin{align} \label{3.16}
\frac{\delta^n \Gamma_{\text{st}} [\phi]}
{\delta \phi (x_1) \dotsm \delta \phi (x_n)}
\end{align}
can alternatively be computed by differentiating the functional $\Gamma [\,\overline{f}=0; \,\chib] = \overline{\Gamma} \,[\chib]$ with respect to the background $\chib$:
\begin{align} \label{3.17}
\frac{\delta^n \Gamma [\,\overline{f}=0; \,\chib]}
{\delta \chib (x_1) \dotsm \delta \chib (x_n)}
\Bigg \rvert_{\chi_{\text{B}}=\phi}
& \equiv
\frac{\delta^n \Gamma_{\text{st}} [\phi]}
{\delta \phi (x_1) \dotsm \delta \phi (x_n)}.
\end{align}
In the case at hand the equality of \eqref{3.16} and \eqref{3.17} is trivial since $\Gamma [\,\overline{f}=0;\,\chib]$ and $\Gamma_{\text{st}} [\chib]$ are exactly equal here.

The situation is less trivial when one applies this formalism to gauge theories, employing a gauge fixing term invariant under background gauge transformations. Then the analogs of the $n$-point functions \eqref{3.16} and \eqref{3.17} are not exactly equal, but they are equal ``on-shell''. As a result, both sets of correlators give rise to the same set of physical $S$-matrix elements \cite{back}. 
The important conclusion is that even then the functional $\overline{\Gamma}$ which obtains by requiring that the fluctuation has no expectation value (\,$\overline{f} =0$) and depends only on one field ($\chib \equiv \phi$) contains all of the physical, gauge--invariant information.

Before continuing let us summarize the status of the various metrics, all conformal to one another, that enter the construction. First, there is the \textbf{reference metric $\boldsymbol{\widehat{g}_{\mu \nu}}$}, a classical field which is fixed once and for all and never gets varied. Second, there is the \textbf{quantum metric}, the integration variable
\begin{align} \label{3.18}
\gamma_{\mu \nu} 
& = \chi^{2 \nu} \, \widehat{g}_{\mu \nu}
= \left( \chib + f \right)^{2 \nu} \, \,\widehat{g}_{\mu \nu}.
\end{align}
In the canonical approach this metric corresponds to an operator. Third, there is the \textbf{background metric} defined by
\begin{align} \label{3.19}
\overline{g}_{\mu \nu} 
& \equiv \chib^{2 \nu} \, \,\widehat{g}_{\mu \nu}.
\end{align}
It is a classical field again which is considered variable, however. In particular it can be adjusted to achieve $\overline{f} =0$ if this is desired. Fourth, there is the \textbf{expectation value of the quantum metric}
\begin{align} \label{3.20}
g_{\mu \nu} & \equiv \langle \gamma_{\mu \nu} \rangle 
\equiv
\left\langle \left( \chib + f \right)^{2 \nu} \right\rangle 
\, \widehat{g}_{\mu \nu}.
\end{align}
And finally, fifth, there is the \textbf{metric with the conformal factor $\boldsymbol{\phi}$}. As $\phi \equiv \chib + \overline{f} = \chib + \langle f \rangle$, it reads
\begin{align} \label{3.21}
\breve{g}_{\mu \nu}
& \equiv
\phi^{2 \nu} \, \,\widehat{g}_{\mu \nu}
\equiv
\bigl( \chib + \left\langle f \right\rangle \bigr)^{2 \nu} \, \,
\widehat{g}_{\mu \nu}.
\end{align}
In general $g_{\mu \nu}$ and $\breve{g}_{\mu \nu}$ are not exactly equal. However, they are approximately equal if the quantum fluctuations of $f$ are small. In $d=4$ where $\nu=1$, for instance, we have
\begin{align} \label{3.22}
\begin{split}
g_{\mu \nu}
& =
\overline{g}_{\mu \nu} 
+ \left[ 2 \, \chib \, \left\langle f \right\rangle
+ \left\langle f^2 \right\rangle
\right] \, \widehat{g}_{\mu \nu}
\\
\breve{g}_{\mu \nu}
& =
\overline{g}_{\mu \nu} 
+ \left[ 2 \, \chib \, \left\langle f \right\rangle
+ \left\langle f \right\rangle^2
\right] \, \widehat{g}_{\mu \nu}.
\end{split}
\end{align}
Hence the difference $g_{\mu \nu} - \breve{g}_{\mu \nu} =
\left[ \langle f^2 \rangle - \langle f \rangle^2 \right] \,
\widehat{g}_{\mu \nu}$ is proportional to the variance of $f$ so that $g_{\mu \nu}$ and $\breve{g}_{\mu \nu}$ are not very different if the fluctuations of $f$ are ``small''. However, in order to make this statement precise one first would have to give a meaning to the expectation value of the operator product $f^2$ with both operators at the same point, something we shall not attempt here. Notice also that $\breve{g}_{\mu \nu}$ reduces to $\overline{g}_{\mu \nu}$ if $\overline{f} =0$ while $g_{\mu \nu}$ does not: $g_{\mu \nu} = \overline{g}_{\mu \nu} + \langle f^2 \rangle \, \widehat{g}_{\mu\nu}$.

The metrics $g_{\mu \nu}$ and $\overline{g}_{\mu \nu}$ are analogous to the fields with the same names in the construction of the exact gravitational average action \cite{mr}. Certain differences arise, however, since there a linear background--quantum split is performed at the level of the full metric, while in the present approach the split is linear at the level of the conformal factor. In ref.\ \cite{mr} where the integral over all metrics $\gamma_{\mu \nu}$ is dealt with, one decomposes $\gamma_{\mu \nu} = \overline{g}_{\mu\nu} + h_{\mu \nu}$ and then integrates over the fluctuation $h_{\mu \nu}$. As a result, $g_{\mu \nu} = \langle \gamma_{\mu \nu} \rangle = \overline{g}_{\mu\nu} + \langle h_{\mu \nu} \rangle$ is linear in the expectation value of the fluctuation so that there is no difference between $g_{\mu \nu}$ and $\breve{g}_{\mu \nu}$. In the present setting, on the other hand, the metric $\gamma_{\mu \nu}$ is parameterized by the fluctuation in a \textit{nonlinear} way: $\gamma_{\mu \nu} = (\chib + f)^{2 \nu} \, \, \widehat{g}_{\mu \nu}$. This nonlinearity is the price we have to pay if we want the CREH action to look like that of a standard scalar $\phi^4$-theory.
\subsection{The Average Action of the Conformal Factor}\label{s3.2}
From the technical point of view, the main problem consists in (approximately) computing the path integral \eqref{3.5}. Next we shall set up an RG formalism which translates this problem into the equivalent problem of solving a certain functional RG equation subject to a boundary condition involving $S$. 

\noindent
\textbf{(A) Introducing a mode cutoff}~Using a variant of the effective average action for scalars \cite{avact,livrev} we modify \eqref{3.5} by introducing a mode--cutoff term into the path integral defining $W$:
\begin{align} \label{3.23}
\begin{split}
& \exp \bigl( W_k [J; \chib] \bigr)
\\
& \phantom{{=}} =
\int \!\! \mathcal{D} f~
\exp \left( - S [\chib +f] - \Delta_k S [f; \chib]
+ \int \!\! \mathrm{d}^d x ~\sqrt{\widehat{g}\,} \,
J (x) \, f (x) \right).
\end{split}
\end{align}
The action $\Delta_k S [f; \chib]$ is to be constructed in such a way that the factor $\exp (- \Delta_k S)$ suppresses the long--wavelength modes of $f (x)$ with momenta $p \lesssim k$ while it does not affect the short--wavelength modes with $p \gtrsim k$. In order to arrive at an FRGE of the familiar second--order type we take $\Delta_k S$ to be quadratic in $f$:
\begin{align} \label{3.24}
\Delta_k S [f; \chib]
& = 
\tfrac{1}{2} \, \int \!\! \mathrm{d}^d x~\sqrt{\widehat{g}\,} \,
f (x) \, \mathcal{R}_k [\chib] \, f (x).
\end{align}
Here $\mathcal{R}_k$ is a pseudodifferential operator which may depend on the background field. Allowing for this $\chib$-dependence is crucial in order to implement ``background independence'' \cite{A,R,T} and to give a ``proper'' meaning to the coarse graining scale $k$ in a theory with a dynamical metric.

\noindent
\textbf{(B) Giving a meaning to $\boldsymbol{k}$}~In flat space the parameter $k$, by elementary Fourier theory, has the interpretation of the inverse length scale over which the microscopic fields are averaged or ``coarse grained''. If we want to have a similar interpretation in quantum gravity we must decide with respect to which metric this length scale is measured. In the background field approach, there is a canonical candidate for a metric measuring the coarse graining scale, namely the background metric $\overline{g}_{\mu \nu} = \chib^{2 \nu} \, \, \widehat{g}_{\mu\nu}$. In fact, as we discussed in the Introduction, the RG flow becomes ``background independent'' (in the sense of \cite{A,R,T}) if $\Delta_k S$ is constructed from $\overline{g}_{\mu \nu}$, or $\chib$ here, rather than from a rigid metric. The key property of $\mathcal{R}_k [\chib]$ is to distinguish ``long--wavelength'' and ``short--\-wave\-length'' modes of $f (x)$ whereby the ``length'' is defined in terms of $\overline{g}_{\mu \nu}$, i.\,e.\ the background conformal factor $\chib$.

The advantage of using the background field method is that at an intermediate stage it decouples the field integrated over, the fluctuation $f$, from the field that fixes the physical value of $k$, namely $\chib$. At the very end, after the quantization has been performed and the RG trajectories are known, we may set $\overline{f} =0$ without loosing information. Then the scale dependent version of the single--argument functional defined above, $\overline{\Gamma}_k [\phi \equiv \chib]$, depends only on one conformal factor, corresponding to ``the'' metric $g_{\mu \nu}$, and its parameter $k$ is a momentum measured, indirectly, with respect to this metric.

\noindent
\textbf{(C) Which spectrum is cut off\,?}~A cutoff operator $\mathcal{R}_k$ with the desired properties can be constructed along the following lines. We think of the functional integral \eqref{3.23} over $f$ as being organized according to eigenfunctions of the Laplace--Beltrami operator constructed from $\overline{g}_{\mu \nu}$:
\begin{align} \label{3.25}
\overline{\Box} & \equiv \overline{g}\,^{-1/2} \,
\partial_\mu \, \overline{g}\,^{1/2} \, 
\overline{g}\,^{\mu \nu} \, \partial_\nu.
\end{align}
Expanding $f$ in terms of $(- \overline{\Box}\,)$-eigenfunctions, the task of $\mathcal{R}_k$ is to suppress those with eigenvalues smaller than $k^2$ by giving them a ``mass'' of the order $k$, while those with larger eigenvalues must remain ``massless'' \cite{avact,livrev}. In the simplest case when the $f$-modes have a kinetic operator proportional to $\overline{\Box}$ itself the rule is that the correct $\mathcal{R}_k$ when added to $\Gamma_k^{(2)}$ leads to the replacement
\begin{align} \label{3.26}
(- \overline{\Box}\,)
& \longrightarrow
(- \overline{\Box} \,) 
+ k^2 \, R^{(0)} \bigl( \tfrac{- \overline{\Box}\,}{k^2\,} \bigr).
\end{align}
Here $R^{(0)} (z)$ is an arbitrary ``shape function'' interpolating between $R^{(0)} (0) =1$ and $R^{(0)} (\infty) =0$, with a transition region centered around $z=1$. These conditions guarantee that the effective inverse propagator of the long-- and short--wavelength modes is $- \overline{\Box} + k^2$ and $- \overline{\Box}$, respectively, and that the long/short--transition is at the $- \overline{\Box}$-eigenvalue $k^2$, as it should be.

The coarse graining scale $\ell = \ell (k)$ corresponding to the cutoff value $k$ is found by investigating the properties of the $-\overline{\Box}$-eigenfunction with eigenvalue $k^2$, the so-called ``cutoff mode'' \cite{jan1,jan2}: one determines its typical scale of variation with respect to $x$ (a period, say) and converts this coordinate length to a physical, i.\,e.\ proper length using $\overline{g}_{\mu \nu}$. The result, $\ell (k)$, is an approximate measure for the extension of the spacetime volumes up to which the dynamics has been ``coarse grained''. If $\overline{g}_{\mu \nu}$ is close to a flat metric, $\ell (k)$ equals approximately $\pi / k$. (See \cite{jan1,jan2} for a detailed discussion.) It is in this sense that the background metric $\overline{g}_{\mu \nu}$, or rather its conformal factor $\chib$, determines the physical (proper) scale of $k$.

Defining the scale $k$ as a cutoff in the spectrum of the covariant Laplacian built from $\overline{g}_{\mu \nu}$ is in accord with the construction of the exact gravitational average action in \cite{mr}; there, too, it is the background metric which sets the scale of $k$.

\noindent
\textbf{(D) Matter fields vs.\ quantized gravity}~While the above choice of $\mathcal{R}_k$ appears very natural, and in fact is the only meaningful one in the gravitational context, every standard quantization and RG scheme which treats $\phi$ as an ordinary scalar uses a differently defined cutoff, namely one based upon $\widehat{\Box}$. Here $\widehat{\Box}$ denotes the Laplace--Beltrami operator pertaining to the reference metric, 
$\widehat{\Box} =\widehat{g}\,^{-1/2} \,
\partial_\mu \, \widehat{g}\,^{1/2} \, 
\widehat{g}\,^{\mu \nu} \, \partial_\nu$, and $\mathcal{R}_k$ is designed to implement the replacement
\begin{align} \label{3.27}
(- \widehat{\Box})
& \longrightarrow
(- \widehat{\Box}) 
+ k^2 \, R^{(0)} \bigl( \tfrac{- \widehat{\Box}}{k^2\,} \bigr).
\end{align}
In this case the proper scale of $k$ is determined by the metric $\widehat{g}_{\mu \nu}$ which, however, at no stage of the construction acquires any physical meaning. As we emphasized, $\widehat{g}_{\mu \nu}$ is never varied. It ``knows'' nothing about the true (``on-shell'') metric of spacetime, namely the particular background metric which adjusts itself dynamically upon setting $\langle f \rangle =0$. The scheme \eqref{3.27} is the correct choice if one considers $\chi$ a standard scalar field on a non-dynamical spacetime with metric $\widehat{g}_{\mu \nu}$, on flat space ($\widehat{g}_{\mu \nu}=\delta_{\mu \nu}$), for instance. The average action formalism based upon \eqref{3.27} reproduces all the familiar results of perturbation theory, the $\ln (k)$-running of the quartic coupling in $\phi^4$-theory, for instance.

Since $\widehat{g}_{\mu \nu}$ is a rigid metric, the flow resulting from the substitution \eqref{3.27} is not ``background independent'' in the sense of \cite{A,R,T}, while \eqref{3.26} does indeed give rise to a ``background independent'' RG flow.

As we shall see, the flow based upon the $\overline{\Box}$-scheme \eqref{3.26} is \textit{extremely} different from the one for standard scalars. The reason is, of course, that via the $\chib$-dependence of $\overline{\Box}$ the gravitational field itself sets the scale of $k$. The difference between \eqref{3.26} and \eqref{3.27} becomes manifest when we recall that the Laplacians of 
$\widehat{g}_{\mu \nu}$ and 
$\overline{g}_{\mu \nu} = \chib^{2 \nu} \, \,\widehat{g}_{\mu \nu}$ are related by
\begin{align} \label{3.28}
\overline{\Box} 
& =
\chib^{-2 \nu} \, \widehat{\Box}
+ \mathcal{O} (\partial \chib).
\end{align}
The factor $\chib^{-2 \nu}$ leads to dramatic modifications of the RG flow whereas the $\mathcal{O} (\partial \chib)$-terms are less important; within the Einstein--Hilbert truncation they play no role.

\noindent
\textbf{(E) Defining $\boldsymbol{\Gamma_k}$}~The remaining steps of the construction follow the familiar rules \cite{avact,avactrev,livrev}. One defines the $k$-dependent field expectation value
\begin{align} \label{3.29}
\overline{f} (x) & \equiv \big\langle f(x) \big\rangle_k
=
\frac{1}{\sqrt{\widehat{g} (x)\,}\,} \,
\frac{\delta W_k [J; \chib]}{\delta J (x)},
\end{align}
solves for the source, $J (x) = J_k [\,\overline{f}; \chib] (x)$, and finally defines the effective average action $\Gamma_k$ as the Legendre transform of $W_k$ with $\Delta_k S [\,\overline{f}; \chib]$ subtracted:
\begin{align} \label{3.30}
\begin{split}
\Gamma_k [\,\overline{f}; \chib]
& =
\int \!\! \mathrm{d}^d x ~\sqrt{\widehat{g}\,} \,\,
\overline{f} (x) \, J_k [\,\overline{f}; \chib] (x)
- W_k \bigl[ J_k [\,\overline{f}; \chib]; \chib \bigr]
\\
& \phantom{{==}}
- \tfrac{1}{2} \, \int \!\! \mathrm{d}^d x ~\sqrt{\widehat{g}\,} \,\,
\overline{f} \, \mathcal{R}_k [\chib] \, \overline{f}.
\end{split}
\end{align}
In analogy with \eqref{3.10} and \eqref{3.11} we also introduce
\begin{gather}
\label{3.31}
\Gamma_k [\phi, \chib]
\equiv
\Gamma_k [\,\overline{f}=\phi-\chib;\,\chib]
\\
\label{3.32}
\overline{\Gamma}_k [\phi]\
\equiv \Gamma_k [\phi, \phi]
= 
\Gamma_k [\,\overline{f}=0;\,\chib=\phi].
\end{gather}

\noindent
\textbf{(F) The flow equation}~The main properties of $\Gamma_k$ are easily established along the same lines as in standard scalar theories \cite{avact,avactrev,livrev}. In particular, differentiating \eqref{3.23} with respect to $k$ leads to the following FRGE which governs the scale dependence of $\Gamma_k$:
\begin{align} \label{3.33}
k \partial_k \, \Gamma_k [\,\overline{f};\chib]
& = \tfrac{1}{2} \, \tr
\left[ \left( \Gamma_k^{(2)} [\,\overline{f};\chib]
+ \mathcal{R}_k [\chib] \right)^{-1} \,
k \partial_k \, \mathcal{R}_k [\chib] \right].
\end{align}
Here $\Gamma_k^{(2)}$ is the matrix of second functional derivatives of $\Gamma_k [\,\overline{f};\chib]$ with respect to $\overline{f}$ at fixed $\chib$. In bra--ket notation,
\begin{align} \label{3.34}
\langle x \lvert \Gamma_k^{(2)} \rvert y \rangle
& =
\frac{1}{\sqrt{\widehat{g} (x)\,} \, \sqrt{\widehat{g} (y)\,}\,}
\, \frac{\delta^2 \Gamma_k [\,\overline{f};\chib]}
{\delta \overline{f} (x) \, \delta \overline{f} (y)}.
\end{align}
Note that the metric appearing in formulas such as \eqref{3.29} or \eqref{3.34} is $\widehat{g}_{\mu \nu}$ (and not $\overline{g}_{\mu \nu}$). Correspondingly $\tr (\cdots) \equiv \int \! \mathrm{d}^d x~
\sqrt{\widehat{g}\,} \, \langle x \lvert (\cdots) \rvert x \rangle$. Notice also that, since the $\overline{f}$-derivatives are to be performed at fixed $\chib$, the FRGE \eqref{3.33} cannot be formulated in terms of the single--argument functional $\overline{\Gamma}_k$ alone. Hence the relevant theory space consists of functionals depending on two fields, $\overline{f}$ and $\chib$, or alternatively $\phi$ and $\chib$.

By construction $\mathcal{R}_k$ vanishes for $k \to 0$. As a consequence, $\Gamma_k$ reduces to the ordinary effective action in this limit:
\begin{align} \label{3.35}
\begin{split}
\Gamma_{k=0} [\,\overline{f}; \chib] 
& = 
\Gamma [\,\overline{f}; \chib] 
\\
\overline{\Gamma}_{k=0} [\phi]
& =
\overline{\Gamma} \,[\phi].
\end{split}
\end{align}
Hence $\Gamma_{k \to 0}$ and $\overline{\Gamma}_{k \to 0}$ satisfy the relations \eqref{3.14} and \eqref{3.15}, respectively. They entail that $\Gamma_{k=0} [\,\overline{f}; \chib]$ actually depends on the sum $\chib + \overline{f}$ only. This is \textit{not} true for $k \neq 0$, the reason being that in general $\Delta_k S [f; \chib]$ depends on $f$ and $\chib$ separately, not only on their sum. In the opposite limit $k \to \infty$, $\Gamma_{k} [\,\overline{f}; \chib]$ approaches $S [\chib + \overline{f}\,]$ plus a computable correction term, see \cite{liouv} for a detailed discussion of this point.
%
%
%
%
%
%
\section{The CREH Truncation}\label{s4}
\subsection{The Ansatz for $\boldsymbol{\Gamma_k}$}\label{s4.1}
In this section we specialize the as to yet exact flow equation \eqref{3.33} for the ``CREH truncation''\footnote{For a different approach to the quantization of conformal fluctuations see \cite{narpad}.}. It involves two approximations:
\begin{enumerate}
\item The usual Einstein--Hilbert truncation.
\item The conformal reduction: only the conformal factor is quantized while all other degrees of freedom contained in the metric as well as the Faddeev--Popov ghost fields are neglected.
\end{enumerate}
To make the presentation as transparent as possible we specialize for $d=4$ in the following.

The truncation ansatz for $\Gamma_{k} [\,\overline{f}; \chib]$ is given by the reduced functional $S_{\text{EH}} [\chib + \overline{f}\,]$ from eq.\ \eqref{2.7} with a $k$-dependent Newton constant $G_k$ and cosmological constant $\Lambda_k$:
\begin{align} \label{4.1}
\begin{split}
\Gamma_{k} [\,\overline{f}; \chib]
& =
- \frac{3}{4 \pi \, G_k} \,
\int \!\! \mathrm{d}^4 x~\sqrt{\widehat{g}\,} \,
\Big\{ - \tfrac{1}{2} \, \left( \chib + \overline{f}\, \right)
\, \widehat{\Box} \, \left( \chib + \overline{f}\, \right)
\\
& \phantom{{==}- \frac{3}{4 \pi \, G_k} \,
\int \!\! \mathrm{d}^4 x~\sqrt{\widehat{g}\,} \,
\Big\{}
+ \tfrac{1}{12} \, \widehat{R} \, \left( \chib + \overline{f}\, \right)^2
- \tfrac{1}{6} \, \Lambda_k \, \left( \chib + \overline{f}\, \right)^4 
\Big\}.
\end{split}
\end{align}
Here $\chib$ and $\overline{f}$ are still arbitrary functions of $x$. Parametrically the average action also depends on the, equally arbitrary, reference metric $\widehat{g}_{\mu \nu}$ with Ricci scalar $\widehat{R}$ and Laplace--Beltrami operator $\widehat{\Box}$. For this action the Hessian \eqref{3.34} has the form
$\langle x \lvert \Gamma_k^{(2)} \rvert y \rangle =
\Gamma_k^{(2)} \, \delta^4 (x-y) / \sqrt{\widehat{g} (x)\,}$ where $\Gamma_k^{(2)}$ is to be interpreted as a differential operator acting on $x$; it reads
\begin{align} \label{4.2}
\Gamma_{k}^{(2)} [\,\overline{f}; \chib]
& =
- \frac{3}{4 \pi \, G_k} \,
\Big\{ - \widehat{\Box}_x
+ \tfrac{1}{6} \, \widehat{R} (x)
- 2\, \Lambda_k \, \bigl( \chib (x) + \overline{f} (x) \bigr)^2 
\Big\}.
\end{align}
We shall come back to this operator shortly.
\subsection{The projected RG Equations}\label{s4.2}
\noindent
\textbf{(A) The strategy}~In order to determine the $\beta$-functions for the running Newton constant $G_k$ and cosmological constant $\Lambda_k$ we proceed as follows. The first step consists in inserting the ansatz into the flow equation, both on its LHS, where we get $k$-derivatives of $G_k$ and $\Lambda_k$, and on its RHS where we are left with the problem of calculating a functional trace involving $\Gamma_k^{(2)}$. It is sufficient to compute this trace in a derivative expansion which retains only those terms which are also present on the LHS of the flow equation, namely those proportional to the monomials $\phi \, \widehat{\Box} \, \phi$, $\widehat{R} \, \phi^2$, and $\phi^4$ where $\phi \equiv \chib + \overline{f}$. If we then equate the coefficients of equal monomials on the LHS and RHS we find the desired RG equations of $G_k$ and $\Lambda_k$.

\noindent
\textbf{(B) The derivative expansion}~Without loosing information this calculation can be performed with a homogeneous background field: $\chib (x) = const \equiv \chib$.
The following two calculations are necessary then in order to ``project out'' the three monomials of interest:

\noindent\textbf{(i)} Evaluation of the functional trace for a flat metric $\widehat{g}_{\mu \nu}=\delta_{\mu \nu}$ and a non-zero, non-constant field $\overline{f} (x)$. Only the term $\overline{f} \, \widehat{\Box} \, \overline{f}$ must be retained. Comparing it to the relevant term of the LHS,
\begin{align} \label{4.3}
k \partial_k \, \Gamma_{k} [\,\overline{f}; \chib]
& =
+ \frac{3}{4 \pi} \, k \partial_k \, \left( \frac{1}{G_k} \right) \,
\int \!\! \mathrm{d}^4 x~\tfrac{1}{2} \, \overline{f} \, \widehat{\Box} \, \overline{f} + \cdots
\end{align}
yields the $\beta$-function of $G_k$.

\noindent\textbf{(ii)} Evaluation of the functional trace for $\overline{f} \equiv 0$ and $\widehat{g}_{\mu \nu}$ arbitrary whereby only the monomials $\chib^4$ and $\widehat{R}\, \chib^2$ are retained. Comparison with the corresponding terms on the LHS,
\begin{align} \label{4.4}
k \partial_k \, \Gamma_{k} [0; \chib]
& =
- \frac{3}{4 \pi} \, 
\int \!\! \mathrm{d}^4 x~\sqrt{\widehat{g}\,} \,
\left\{ \tfrac{1}{12} \, k \partial_k \, \left( \frac{1}{G_k} \right) \, \widehat{R} \, \chib^2
- \tfrac{1}{6} \, k \partial_k \, \left( \frac{\Lambda_k}{G_k} \right) \, \chib^4 + \cdots \right\}
\end{align}
allows for the computation of $\partial_k \left( \Lambda_k / G_k \right)$ and an alternative determination of $\partial_k G_k$.

Since both the $\phi \, \widehat{\Box} \, \phi$ and the $\widehat{R} \, \phi^2$ term appear with the same prefactor $1 / G_k$ we can derive a $\beta$-function for $G_k$ from either of them. They involve the anomalous dimension $\etan$, and the two versions stemming from the kinetic and the potential term $\propto \phi^2$ will be denoted $\etan^{\text{(kin)}}$ and $\etan^{\text{(pot)}}$, respectively. We do not expect these $\beta$-functions or anomalous dimensions to be exactly equal, but if our approximation makes sense they should be similar at least.

\noindent
\textbf{(C) The explicit form of $\boldsymbol{\mathcal{R}_k}$}~Before we can  embark on these calculations we must address the question of how $\mathcal{R}_k$ is to be adjusted. The IR cutoff at $k$ must be imposed on the spectrum of $\overline{\Box}$, not that of $\widehat{\Box}$. Since $\chib = const$ in the case at hand, the two operators are related by
\begin{align} \label{4.5}
\widehat{\Box} & = \chib^2 \, \overline{\Box}
\end{align}
so that we may reexpress $\Gamma_k^{(2)}$ as
\begin{align} \label{4.6}
\Gamma_{k}^{(2)} [\,\overline{f}; \chib]
& =
- \frac{3}{4 \pi \, G_k} \,
\Big\{ - \chib^2 \, \overline{\Box}
+ \tfrac{1}{6} \, \widehat{R}
- 2\, \Lambda_k \, \left( \chib + \overline{f}\, \right)^2 
\Big\}.
\end{align}
Now we define $\mathcal{R}_k$ in such a way that it leads to the replacement \eqref{3.26} when added to $\Gamma_k^{(2)}$:
\begin{align} \label{4.7}
\begin{split}
& \Gamma_{k}^{(2)} [\,\overline{f}; \chib] + \mathcal{R}_k [\chib]
\\
& \phantom{{=}} =
- \frac{3}{4 \pi \, G_k} \,
\Big\{ \chib^2 \, \bigl[ - \overline{\Box}
+ k^2 \, R^{(0)} (- \overline{\Box} / k^2) \bigr]
+ \tfrac{1}{6} \, \widehat{R}
- 2\, \Lambda_k \, \left( \chib + \overline{f}\, \right)^2 
\Big\}.
\end{split}
\end{align}
As a consequence, the cutoff operator has an explicit dependence on the background field:
\begin{align} \label{4.8}
\begin{split}
\mathcal{R}_k [\chib]
& =
- \frac{3}{4 \pi \, G_k} \, \chib^2 \, k^2 \, 
R^{(0)} \bigl(- \tfrac{\overline{\Box}\,}{k^2\,} \bigr)
\\
& =
- \frac{3}{4 \pi \, G_k} \, \chib^2 \, k^2 \, 
R^{(0)} \bigl(- \tfrac{\widehat{\Box}\,}{\chib^2 \, k^2\,} \bigr)
\end{split}
\end{align}
The two factors of $\chib^2$ appearing in the second line of \eqref{4.8} are the crucial difference between our treatment of the conformal factor and a standard scalar. If, instead of \eqref{3.26}, we had applied the ``substitution rule'' \eqref{3.27} they would have been absent.

\noindent
\textbf{(D) The functional trace}~Upon inserting the above $\mathcal{R}_k$ and reexpressing $\overline{\Box}$ as $\widehat{\Box} / \chib^2$ the flow equation assumes the form
\begin{align} \label{4.9}
\begin{split}
& k \partial_k \, \Gamma_{k} [\,\overline{f}; \chib]
\\
& \phantom{{=}} = \chib^2 \, k^2 \,
\tr \Biggl[
\left\{
\left( 1 - \tfrac{1}{2} \, \etan \right) \,
R^{(0)} \bigl( - \tfrac{\widehat{\Box}\,}{\chib^2 \, k^2\,} \bigr)
- \Bigl( - \frac{\widehat{\Box}}{\chib^2 \, k^2\,} \Bigr) \,
{R^{(0)}}' \bigl( - \tfrac{\widehat{\Box}\,}{\chib^2 \, k^2\,} \bigr)
\right\}
\\
& \phantom{{===}\chib^2 \, k^2 \,\tr \Biggl[} \times
\biggl( - \widehat{\Box} + \tfrac{1}{6} \, \widehat{R}
+ \chib^2 \, k^2 \, R^{(0)} \bigl( - \tfrac{\widehat{\Box}\,}{\chib^2 \, k^2\,} \bigr)
- 2\, \Lambda_k \, \left( \chib + \overline{f}\, \right)^2 
\biggr)^{-1}
\Biggr].
\end{split}
\end{align}
\begin{align} \label{4.9}
\begin{split}
& k \partial_k \, \Gamma_{k} [\,\overline{f}; \chib]
\\
& \phantom{{=}} = \chib^2 \, k^2 \,
\tr \Biggl[
\Bigg\{
\left( 1 - \frac{1}{2} \, \etan \right) \,
R^{(0)} \biggl( - \frac{\widehat{\Box}\,}{\chib^2 \, k^2\,} \biggr)
- \biggl( - \frac{\widehat{\Box}}{\chib^2 \, k^2\,} \biggr) \,
{R^{(0)}}' \biggl( - \frac{\widehat{\Box}\,}{\chib^2 \, k^2\,} \biggr)
\Bigg\}
\\
& \phantom{{===}\chib^2 \, k^2 \,\tr \Biggl[} \times
\Biggl( - \widehat{\Box} + \frac{1}{6} \, \widehat{R}
+ \chib^2 \, k^2 \, R^{(0)} \biggl( - \frac{\widehat{\Box}\,}{\chib^2 \, k^2\,} \biggr)
- 2\, \Lambda_k \, \left( \chib + \overline{f}\, \right)^2 
\Biggr)^{-1}
\Biggr].
\end{split}
\end{align}
In evaluating the derivative $\partial_k \mathcal{R}_k$ we encountered the anomalous dimension $\etan$, defined in the same way as in \cite{mr}:
\begin{align} \label{4.10}
\etan & \equiv + k \partial_k \, \ln G_k.
\end{align}

Note that in eq.\ \eqref{4.9} the overall minus sign of $\mathcal{R}_k$, and hence $k \partial_k \mathcal{R}_k$, got canceled against the overall minus sign of $\Gamma_k^{(2)} + \mathcal{R}_k$ in \eqref{4.7}. This is the step where, within the present setting, the transition from the ``original'' to the ``inverted'' picture has taken place. The factor $(\cdots)^{-1}$ under the trace of \eqref{4.9} is the propagator of a mode with positive kinetic, but negative potential energy. This is an example of the ``$\mathcal{Z}_k = z_k$ rule'' discussed in \cite{mr} and \cite{oliver2}.

The only specialization which entered eq.\ \eqref{4.9} is $\chib = const$; the reference metric $\widehat{g}_{\mu \nu}$ and the fluctuation average $\overline{f}$ are still arbitrary. Therefore equation \eqref{4.9} can serve as the starting point for the two calculations (i) and (ii) which must be performed at this point. 

\noindent
\textbf{(E) The resulting beta functions}~The details of the calculations, for an arbitrary shape function $R^{(0)}$ and any spacetime dimension $d$ can be found in \cite{creh1}. Here we only present the final result for the RG equations, in $d=4$ dimensions, employing the ``optimized'' shape function of ref.\ \cite{opt}:
\begin{align} \label{4.29}
R^{(0)} (z) = \left( 1-z \right) \, \theta (1-z).
\end{align}
It is convenient to express the coupled system of differential equations for Newton's constant and the cosmological constant in terms of the dimensionless couplings
\begin{align} \label{4.19}
g_k \equiv k^2 \, G_k, \qquad 
\lambda_k \equiv \Lambda_k / k^2.
\end{align}
This choice of variables makes the system autonomous:
\begin{align}
\label{4.20}
k \partial_k \, g_k & = \beta_g (g_k, \lambda_k)
= \bigl[ 2 + \etan (g_k, \lambda_k) \bigr] \, g_k
\\
\label{4.21}
k \partial_k \, \lambda_k & = \beta_\lambda (g_k, \lambda_k).
\end{align}
For the anomalous dimension coming from the kinetic term we obtain explicitly
\begin{align} \label{4.35}
\etan^{\text{(kin)}} (g_k, \lambda_k)
& =
- \frac{2}{3 \pi} \, 
\frac{g_k \,\lambda_k^2}{\left( 1 - 2 \lambda_k \right)^4\,}.
\end{align}
The one derived from the potential has the familiar structure \cite{mr}
\begin{align} \label{4.22}
\etan^{\text{(pot)}} (g_k, \lambda_k)
& =
\frac{g_k \, B_1 (\lambda_k)}{1 - g_k \, B_2 (\lambda_k)}
\end{align}
with the following $B$-functions:
\begin{subequations} \label{4.36}
\begin{align}
\label{4.36a}
B_1 (\lambda_k)
& =
\frac{1}{3 \pi} \, \left( \frac{1}{4} - \lambda_k \right) \,
\frac{1}{\left( 1 - 2 \lambda_k \right)^2\,}
\\
\label{4.36b}
B_2 (\lambda_k)
& =
- \frac{1}{12 \pi} \, \left( \frac{1}{3} - \lambda_k \right) \,
\frac{1}{\left( 1 - 2 \lambda_k \right)^2\,}.
\end{align}
\end{subequations}
For $\beta_\lambda$ one finds
\begin{align} \label{4.37}
\beta_\lambda (g_k, \lambda_k)
& =
- \left( 2-\etan \right) \, \lambda_k
+ \frac{g_k}{4 \pi} \, \left( 1 - \frac{1}{6} \, \etan \right) \,
\frac{1}{1 - 2 \lambda_k}
\end{align}
where either $\etan^{\text{(kin)}}$ or $\etan^{\text{(pot)}}$ is to be inserted for $\etan$.

Using $\etan^{\text{(kin)}}$ the $\beta$-functions have poles at $\lambda = 1/2$ and are regular otherwise. The physically relevant part of the parameter space is the half plane to the left of this line ($\lambda < 1/2$), as in the full theory \cite{frank1}.

With $\etan^{\text{(pot)}}$ the boundary of the ``physical'' parameter space is given by a curve to the left of the $\lambda=1/2$--line. Along this line, $1-g\, B_2 (\lambda) =0$, so that $\etan^{\text{(pot)}}$ diverges there, $\lvert \etan^{\text{(pot)}} \rvert = \infty$. Parameterizing this curve as $g = g_\eta^{\text{(pot)}} (\lambda)$ we have explicitly
\begin{align} \label{4.38}
g_\eta^{\text{(pot)}} (\lambda)
& =
12 \pi \, \frac{\left( 1- 2\lambda\right)^2}{\lambda - 1/3}.
\end{align}

In either of the two cases, the existence of a boundary in 
$(g, \lambda)$-space entails that some of the RG trajectories terminate already at a finite value of $k$ when they run into the boundary line. Within the full Einstein--Hilbert truncation, the status of the singularities has been discussed in detail in the literature \cite{frank1,frank2,h3,litimgrav}.
They have been interpreted as a breakdown of the truncation in the infrared. The continuation to $k=0$ would presumably require a more general ansatz for $\Gamma_k$.
\subsection{Comparison with the Standard Scalar FRGE}\label{s4.4}
One might wonder how the RG equations for the conformal factor relate to those for a standard scalar \cite{avactrev}. The comparison reveals that both the structure of the equation and their solutions are quite different in the two cases. We shall see this in more detail in Section \ref{s5}. Here we only mention the most striking deviation.

Let us consider an RG trajectory in a regime where the anomalous dimension is small so that we may approximate $\etan =0$. (In the next section we shall see that there are indeed trajectories with $\etan \approx 0$ over a large range of scales.) Then \eqref{4.10} integrates to $G_k = const \equiv \overline{G}$, and the equation for $\lambda_k$ involves the correspondingly simplified $\beta$-function \eqref{4.37}, with $g_k \equiv \overline{G} \, k^2$. In terms of the ordinary, dimensionful cosmological constant $\Lambda_k \equiv k^2 \, \lambda_k$ this RG equation reads
\begin{align} \label{4.40}
k \partial_k \, \Lambda_k
& =
\frac{\overline{G}}{4 \pi} \, \frac{k^6}{k^2 - 2 \Lambda_k}.
\end{align}
In particular, when $\Lambda_k \ll k^2$ it simplifies to
\begin{align} \label{4.41}
k \partial_k \, \Lambda_k
& =
\frac{1}{4 \pi} \, \overline{G} \, k^4.
\end{align}
Obviously \textit{the RG equations of the CREH truncation imply a quartic running of the cosmological constant as long as $\mathit{\Lambda}$ is small and $\mathit{G}$ is approximately constant.}

With quantum gravity in the back of our mind this result is no surprise. It is exactly what one finds in the full Einstein--Hilbert truncation \cite{mr}, except for the prefactor of $\overline{G} \, k^4$ which is anyhow non-universal. In fact, the $k^4$-running \eqref{4.41} is what all methods for summing zero--point energies would agree upon. In particular it can be seen as a reflection of the well known quartic divergences which appear in all Feynman diagram calculations (and are usually ``renormalized away''). So there can be no doubt that \eqref{4.41} is the physically correct answer for the regime considered.

On the other hand, from the scalar field perspective, the quartic running \textit{is} a surprise. In the CREH ansatz for $\Gamma_k$ the cosmological constant $\Lambda_k$ plays the role of a $\phi^4$-coupling constant which behaves as $\Lambda_k \propto k^4$ here. This very strong scale dependence has to be contrasted with the much weaker, merely logarithmic $k$-dependence one finds in an ordinary scalar theory on a $4$-dimensional flat spacetime (provided $k$ is above all mass thresholds, if any).

The origin of this significant difference in the RG running of the $\phi^4$-coupling, proportional to $\ln (k)$ for a standard scalar and $\propto k^4$ for the conformal factor, is clear: The conformal factor determines the proper scale of the cutoff, while a scalar matter field does not.
When we constructed the operator $\mathcal{R}_k$ in Subsection \ref{s4.2} we explained how the special status of the conformal factor comes into play. We saw that if the coarse graining scale is to be given a physical meaning, $k$ should be a cutoff in the spectrum of the \textit{background field dependent} operator $\overline{\Box}$, and this led to the substitution rule \eqref{3.26}.

Thus it becomes obvious that ``background independence'' leads to RG equations different from those of a scalar matter field. In the construction of the exact gravitational average action in \cite{mr} where all degrees of freedom carried by the metric are quantized ``background independence'' has likewise been taken care of. There it is the full background metric $\overline{g}_{\mu\nu}$, the generalization of $\chib^2$ here, which enters $\mathcal{R}_k$ and sets the scale of $k$.
%
%
%
\section{Asymptotic Safety in the CREH Truncation}\label{s5}
In this section we analyze the physical contents of the RG flow in the CREH truncation, being particularly interested in the asymptotic safety issue.
\subsection{Antiscreening}\label{s5.1}
From the definition \eqref{4.10} it follows that the RG running of the dimensionful Newton constant is given by
\begin{align} \label{5.1}
k \partial_k \, G_k & = \etan \, G_k.
\end{align}
If $\etan >0$, Newton's constant increases with increasing mass scale $k$, while it decreases if $\etan <0$. In analogy with gauge theory one refers to the first case as ``screening'', the second as ``antiscreening''. In the full\footnote{Here and in the following ``full calculation'' always refers to the complete calculation within the Einstein--Hilbert truncation in ref.\ \cite{mr}.} calculation, $\etan$ was of the antiscreening type in the entire physical part of the $(g, \lambda)$-plane.

If we determine \textbf{$\boldsymbol{\etan}$ from the kinetic term}, the corresponding CREH result $\etan^{\text{(kin)}}$ is negative for any value of $g>0$ and $\lambda$. This corresponds to the antiscreening case: Newton's constant decreases at high energies. So the remarkable result is that the quantization of the conformal factor alone is already sufficient to obtain gravitational antiscreening. The spin-2 character of the metric field seems not essential and its selfinteractions (vertices) coming from $\int \! \mathrm{d}^4 x \sqrt{g\,} \, R$ seem not to play the dominant role. 

If we take the \textbf{$\boldsymbol{\etan}$ from the potential} we find that, if $g>0$,
\begin{align} \label{5.3}
\begin{split}
\etan^{\text{(pot)}} & \leq 0,
\quad \text{if } \lambda \geq 1/4, \\
\etan^{\text{(pot)}} & >0,
\quad \text{if } \lambda < 1/4.
\end{split}
\end{align}
The anomalous dimension $\etan^{\text{(pot)}}$ vanishes along the line $\lambda = 1/4$.
\subsection{Fixed Points}\label{s5.2}
Next we search for fixed points of the system of differential equations \eqref{4.20}, \eqref{4.21}, i.\,e.\ points $(g_\ast, \lambda_\ast)$ such that $\beta_g (g_\ast, \lambda_\ast) = 0 = \beta_\lambda (g_\ast, \lambda_\ast)$.

From \eqref{4.35}, \eqref{4.36}, and \eqref{4.37} it is obvious that for either choice of $\etan$ the system has a fixed point at the origin, referred to as the Gaussian Fixed Point (GFP): $g_\ast^{\text{GFP}} = \lambda_\ast^{\text{GFP}} = 0$.

A Non-Gaussian Fixed Point (NGFP), if any, would satisfy the condition $\beta_g =0$ with non-zero values of $g_\ast$ or $\lambda_\ast$ such that $\etan (g_\ast, \lambda_\ast) = -2$. Upon inserting $\etan=-2$ into \eqref{4.37} the condition $\beta_\lambda =0$ assumes the simple form
\begin{align} \label{5.4}
g_\ast & = 12 \pi \, \lambda_\ast \, \left( 1 - 2 \lambda_\ast
\right).
\end{align}
The second condition for $g_\ast$ and $\lambda_\ast$ depends on the choice for $\etan$.

If we use the \textbf{$\boldsymbol{\etan}$ from the kinetic term} given by eq.\ \eqref{4.35} the condition \linebreak$\etan^{\text{(kin)}} (g_\ast, \lambda_\ast)= -2$ reads
\begin{align} \label{5.5}
g_\ast \, \frac{\lambda_\ast^2}{\left( 1 - 2 \lambda_\ast \right)^4\,}
& = 3 \pi.
\end{align}
The system of equations \eqref{5.4}, \eqref{5.5} is easily decoupled by inserting $g_\ast$ of \eqref{5.4} into \eqref{5.5}. Remarkably, one does indeed find a real solution:
\begin{subequations} \label{5.6}
\begin{align}
\label{5.6a}
\lambda_\ast 
& = \frac{1}{2} \, \frac{2^{1/3}}{\left( 1 + 2^{1/3} \right)}
\approx 0.279
\\
\label{5.6b}
g_\ast 
& = 6 \pi \, \frac{2^{1/3}}{\left( 1 + 2^{1/3} \right)^2\,}
\approx 4.650
\end{align}
\end{subequations}
The existence of this NGFP comes as a true surprise; it has no counterpart in ordinary $4$-dimensional $\phi^4$-theory.

If instead we use the \textbf{$\boldsymbol{\etan}$ from the potential} given by \eqref{4.22} with \eqref{4.36} the condition $\etan^{\text{(pot)}} (g_\ast, \lambda_\ast) = -2$ can be written as
\begin{align} \label{5.7}
g_\ast \, \left( \lambda_\ast - \tfrac{5}{18} \right)
& = 4 \pi \, \left( 1 - 2 \lambda_\ast \right)^2.
\end{align}
The coupled equations \eqref{5.7} and \eqref{5.4} can be solved analytically again and they, too, give rise to real and positive fixed point coordinates:
\begin{subequations} \label{5.8}
\begin{align}
\label{5.8a}
\lambda_\ast 
& = \frac{7}{36} \, \left( \sqrt{481/49\,} -1 \right)
\approx 0.415
\\
\label{5.8b}
g_\ast 
& = 12 \pi \, \lambda_\ast \, \left( 1 - 2 \lambda_\ast \right)
\approx 2.665
\end{align}
\end{subequations}

The individual values of $g_\ast$ and $\lambda_\ast$ as obtained from the two calculational schemes do not quite agree. However, this does not come unexpected. The mere coordinates of the fixed point are not directly related to anything observable and, in fact, are scheme dependent or ``non-universal''. On the other hand, the product $g_\ast \lambda_\ast$ has been argued \cite{oliver1,oliver2} to be universal and can be measured in principle. And indeed, the products of the numbers in \eqref{5.6} and in \eqref{5.8} agree almost perfectly within the precision one could reasonably expect:
\begin{align} \label{5.9}
(g_\ast \lambda_\ast)^{\text{(kin)}} \approx 1.296,
\qquad
(g_\ast \lambda_\ast)^{\text{(pot)}} \approx 1.106
\end{align}

According to both calculations the respective NGFP is always located within the physical part of the $(g, \lambda)$-plane.

It is straightforward to generalize the calculations for arbitrary dimensionalities $d$, see ref.\ \cite{creh1}. The numerical results for $d$ between $3$ and $10$ are displayed in Table \ref{tab1}. 
\begin{table}
\begin{center}
\begin{tabular}{|r||c|c|c|c|c|c|}
\hline
$\boldsymbol{d}$ & \textbf{Trunc.} & $\boldsymbol{g_\ast}$ & $\boldsymbol{\lambda_\ast}$ 
& $\boldsymbol{\tau_d}$ & $\boldsymbol{\theta'}$ & $\boldsymbol{\theta''}$ \\
\hline
\hline
3 & full EH & 0.202139 & 0.0651806 & 0.00266329 & 1.11664 & 0.827598 \\
\hline
& CREH, pot & 0.172872 & 0.233092 & 0.00696588 & -3.54754 & 4.92795 \\
\hline
& CREH, kin & 0.391798 & 0.126945 & 0.0194868 & 2.04572 & 3.60445 \\
\hline
\hline
4 & full EH & 0.707321 & 0.193201 & 0.136655 & 1.4753 & 3.04321 \\
\hline
& CREH, pot & 2.6654 & 0.41477 & 1.10553 & 1.47122 & 9.30442 \\
\hline
& CREH, kin & 4.65005 & 0.278753 & 1.29622 & 4.0 & 6.1837 \\
\hline
\hline
5 & full EH & 2.85863 & 0.234757 & 0.472851 & 2.76008 & 5.12941 \\
\hline
& CREH, pot & 26.9696 & 0.557727 & 5.01577 & 5.81627 & 12.0556 \\
\hline
& CREH, kin & 42.3258 & 0.417188 & 5.06681 & 6.27681 & 8.6899 \\
\hline
\hline
6 & full EH & 13.8555 & 0.255477 & 0.950958 & 4.48592 & 7.07967 \\
\hline
& CREH, pot & 243.547 & 0.674559 & 10.5272 & 10.8493 & 14.3777 \\
\hline
& CREH, kin & 361.57 & 0.537523 & 10.221 & 8.81712 & 11.1844 \\
\hline
\hline
7 & full EH & 76.3589 & 0.269073 & 1.5241 & 6.51007 & 8.9431 \\
\hline
& CREH, pot & 2134.67 & 0.77282 & 16.5886 & 17.0223 & 15.9635 \\
\hline
& CREH, kin & 3069.3 & 0.641211 & 15.9154 & 11.591 & 13.6754 \\
\hline
\hline
8 & full EH & 464.662 & 0.279376 & 2.16389 & 8.78536 & 10.7446 \\
\hline
& CREH, pot & 18744.8 & 0.857143 & 22.7691 & 24.6444 & 16.0092 \\
\hline
& CREH, kin & 26451.9 & 0.730796 & 21.7745 & 14.5789 & 16.1597 \\
\hline
\hline
9 & full EH & 3066.23 & 0.287851 & 2.85326 & 11.2969 & 12.4932 \\
\hline
& CREH, pot & 167205.0 & 0.930559 & 28.9135 & 33.9881 & 12.4239 \\
\hline
& CREH, kin & 233516.0 & 0.808694 & 27.6432 & 17.7666 & 18.6307 \\
\hline
\hline
10 & full EH & 21673.5 & 0.295179 & 3.58153 & 14.044 & 14.1871 \\
\hline
& CREH, pot & 1.52489$ \cdot 10^6$ & 0.995177 & 34.9712 & & \\
\hline
& CREH, kin & 2.11943$ \cdot 10^6$ & 0.876935 & 33.4597 & 21.1433 & 21.0813 \\
\hline
\end{tabular}
\end{center}
\caption{}
\label{tab1}
\end{table}
In all dimensions considered a NGFP is found to exist, with $g_\ast > 0$ and $\lambda_\ast >0$. For each value of $d$, the table contains the results obtained within the full Einstein--Hilbert (EH) truncation as well as the ``pot'' and ``kin'' variants of the CREH truncation. It contains also the generalization of $g_\ast \lambda_\ast$, namely $\tau_d \equiv \lambda_\ast g_\ast^{2/(d-2)}$ which is the fixed point value of the dimensionless combination
$\Lambda_k G_k^{2/(d-2)} = \lambda_k g_k^{2/(d-2)}$. (Note that in $d$ dimensions $g_k = k^{d-2} \, G_k$ and $\lambda_k = k^{-2} \, \Lambda_k$.) It is impressive to see how well the ``pot'' and ``kin'' values of $\tau_d$ agree for any $d \geq 4$. As compared to the full Einstein--Hilbert result, the $\tau_d$-values are always larger by about a factor of 10.

We interpret this factor as indicating that the conformal factor is not the only degree of freedom driving the formation of a NGFP, but its contribution is typical in the sense that it leads to an RG flow which is qualitatively similar to the full one.
\subsection{Critical Exponents of the NGFP}\label{s5.3}
The properties of the RG flow on $(g, \lambda)$-space linearized about the NGFP are determined by the stability matrix
\begin{align} \label{5.10}
B & =
\begin{bmatrix}
\frac{\partial \beta_\lambda}{\partial \lambda} & 
\frac{\partial \beta_\lambda}{\partial g} \\
\frac{\partial \beta_g}{\partial \lambda} & 
\frac{\partial \beta_g}{\partial g}
\end{bmatrix}
\end{align}
evaluated at $(g_\ast, \lambda_\ast)$. Using the same notation as in \cite{oliver1,frank1} we write the corresponding eigenvalue problem as $B \, V = - \theta \, V$ and refer to the negative eigenvalues $\theta$ as the ``critical exponents''. In general $B$ is not expected to be symmetric.

Employing the \textbf{$\boldsymbol{\etan}$ from the kinetic term} the resulting eigenvalues are non-zero and complex. The two critical exponents $\theta_{1,2} = \theta' \pm i \theta''$ form a complex conjugate pair with real and imaginary parts given by, respectively,
\begin{align} \label{5.12}
& & & & & & 
\theta' & = 4,
& \theta'' & = 2 \, \sqrt{2\,} \, \sqrt{1 + 3 \cdot 2^{1/3}\,}
\approx 6.1837
& & & & & & 
\end{align}
The positive real part indicates that the NGFP is UV attractive (attractive for $k \to \infty$) in both directions of the $(g, \lambda)$-plane. The non-vanishing imaginary part implies that near the NGFP the RG trajectories are spirals. This is exactly the same pattern as in the full Einstein--Hilbert truncation \cite{frank1}.

Using instead the \textbf{$\boldsymbol{\etan}$ from the potential} we find the same qualitative behavior, but the exponents are somewhat different:
\begin{align} \label{5.13}
& & & & & & 
\theta' & \approx 1.471, 
& \theta'' & \approx 9.304
& & & & & & 
\end{align}

The discrepancy between \eqref{5.12} and \eqref{5.13} can serve as a rough measure for the accuracy of the calculation. First of all it is gratifying to see that both calculations lead to the same qualitative behavior: attractivity in both directions of parameter space, and a non-zero imaginary part. Numerically, the values for $\theta'$ and $\theta''$ probably can be trusted only within a factor of $2$ or so. In Table \ref{tab1} we display the critical exponents also for the other dimensions and compare them to the values in the full calculation. 

It has to be emphasized, however, that even in an exact treatment of the conformally reduced theory the resulting critical exponents would have no reason to agree with those from full QEG which quantizes also the other degrees of freedom contained in the metric. The field contents of the two theories is different, and so one would expect them to belong to different universality classes, with different $\theta$'s.
\subsection{The Phase Portrait}\label{s5.4}
\begin{figure}[t]
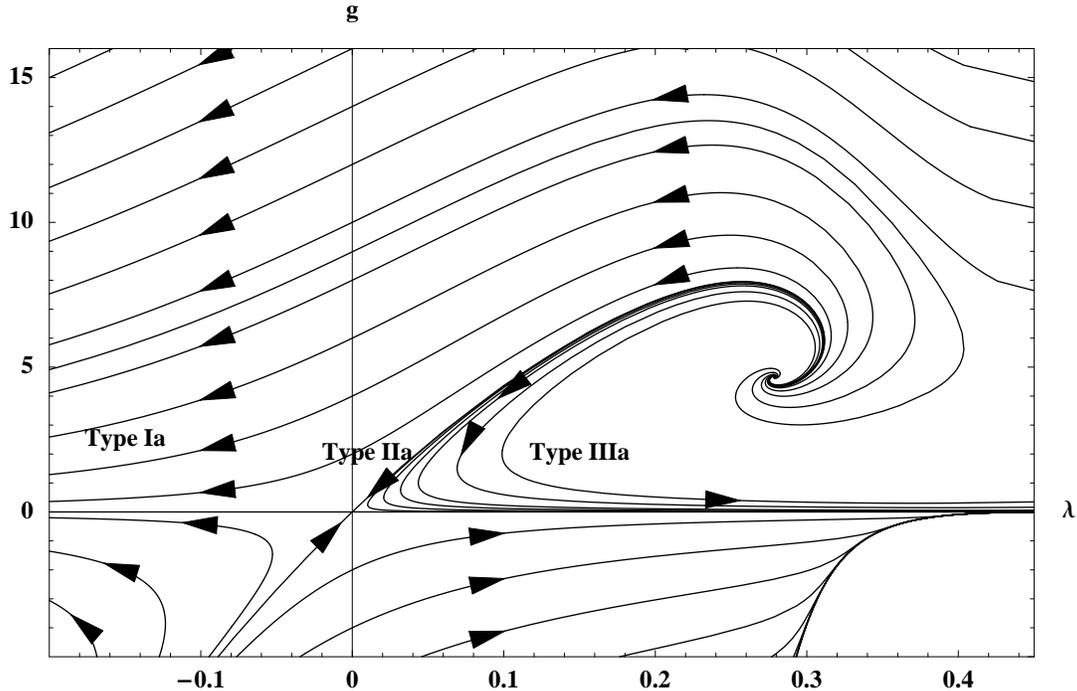

\centering
\pgfuseimage{1a}
\caption{The figure shows the RG flow on the $(g, \lambda)$-plane which is obtained from the CREH truncation with $\etan^{\text{(kin)}}$. The arrows point in the direction of decreasing $k$.}
\label{fig1}
\end{figure}

Finally we solve the coupled equations \eqref{4.20}, \eqref{4.21} numerically in order to obtain the phase portrait of the CREH flow. Using the anomalous dimension $\etan^{\text{(kin)}}$ we find the result displayed in Fig.\ \ref{fig1}. 
This flow diagram is strikingly similar to the corresponding diagram of the full Einstein--Hilbert truncation\footnote{See the diagram in Fig.\ 12 of ref.\ \cite{frank1}.}. The flow is dominated by the NGFP and the GFP at the origin, and we can distinguish three types of trajectories spiraling out of the NGFP. They are heading for negative, vanishing, and positive cosmological constant, respectively, and correspond exactly to the Type Ia, IIa, and IIIa trajectories of the full flow \cite{frank1}. The trajectories of the CREH Types Ia and IIa extend down to $k=0$, those of Type IIIa terminate at a non-zero $k_{\text{term}}$ when they reach $\lambda=1/2$, exactly as in the full theory.

\begin{figure}[t]
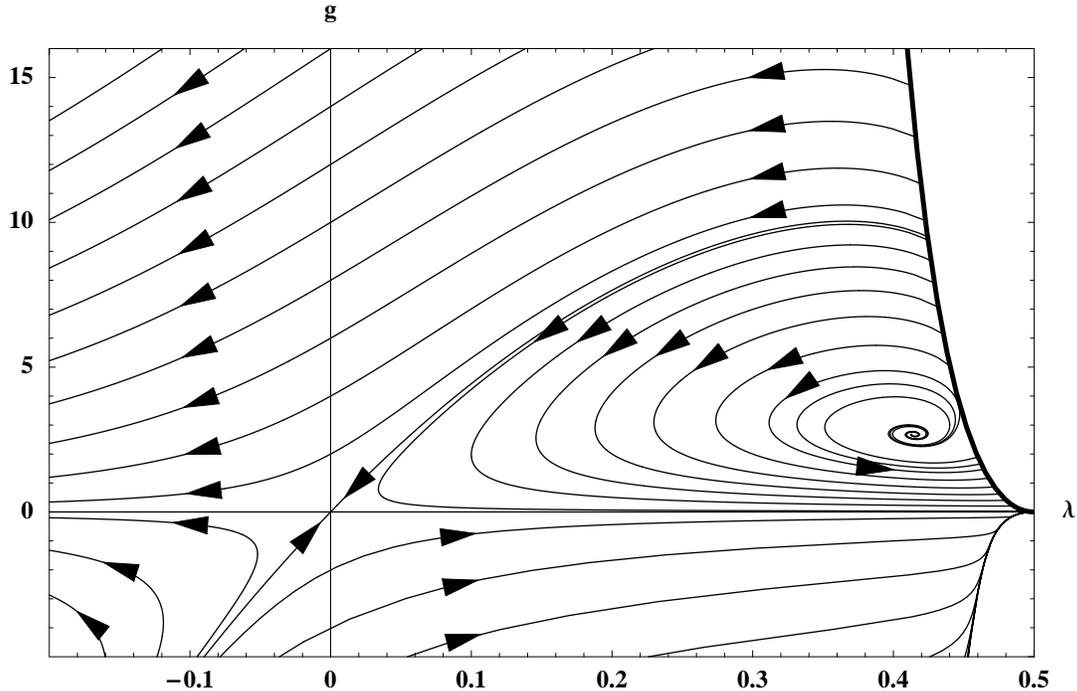

\centering
\pgfuseimage{2a}
\caption{As in Fig.\ \ref{fig1}, but with $\etan^{\text{(pot)}}$. The fat line is the boundary of the physical parameter space on which $\etan^{\text{(pot)}}$ diverges.}
\label{fig2}
\end{figure}

If we solve the RG equations with the second version of the anomalous dimension, $\etan^{\text{(pot)}}$, we obtain the phase portrait shown in Fig.\ \ref{fig2}.
In the vicinity of the GFP and NGFP, respectively, the structure of the flow, again, is exactly the same as in the full theory. The only new feature here is that there exist trajectories which begin and end on the boundary of the physical part of $(g, \lambda)$-space which is given by eq.\ \eqref{4.38}. Even though this feature is different from the full EH flow we see that the conformal factor drives the flow in the same direction as the full metric and is in this sense representative. It is, however, too weak to push the trajectories sufficiently strongly away from the hyperbolic shape they have in absence of any non-trivial RG effects \cite{h3}.

The overall conclusion of this anaysis is that \textit{the RG flow implied by the scalar-like CREH theory, at least in a neighborhood of the two fixed points, is qualitatively identical to that of the full Einstein--Hilbert truncation}. In particular both versions of $\etan$ agree on the existence of a NGFP with precisely the properties required for asymptotic safety.
%
%
%
\section{Summary}\label{s6}
The ultimate theory of quantum gravity we are aiming at should be able to explain rather than merely postulate the spacetime we are living in. Therefore the conceptual foundations of this theory, at no point, should depend on any special non-dynamical spacetime. For QEG this entails that the quantization must not involve any distinguished metric, that is, it should be performed in a ``background independent'' way. Within the asymptotic safety program one tries to define QEG in terms of an RG trajectory on the theory space of the gravitational average action $\Gamma_k$, and this trajectory is supposed to possess an ultraviolet fixed point. Since renormalization group concepts are crucial in this context one is led to ask what is the significance and role of ``background independence'' for RG flows.

Among all the fields we use in order to describe Nature the metric enjoys a special status since it fixes the proper value of any dimensionful physical quantity. When one applies the Kadanoff--Wilson interpretation of RG flows as a sequence of consecutive coarse graining steps to quantum gravity one would like to give an, at least approximate, physical meaning to the notion of a coarse graining scale. The effective average action $\Gamma_k [g_{\mu\nu}, \overline{g}_{\mu \nu}]$ meets this requirement by introducing $k$ as a cutoff in the spectrum of the covariant Laplacian pertaining to the background metric $\overline{g}_{\mu \nu}$. Hence the mass scale $k$ is ``proper'', in the sense of the ``cutoff modes'' \cite{jan1,jan2}, with respect to $\overline{g}_{\mu \nu}$. Therefore the mode suppression term $\Delta_k S$ and, as a result, the cutoff operator $\mathcal{R}_k$ depend on the background field in a non-trivial way which has a strong impact on the resulting RG flow. We saw that to some extent asymptotic safety, the formation of a non-Gaussian fixed point, is an essentially ``kinematical'' phenomenon resulting from this $\overline{g}_{\mu \nu}$-dependence of the cutoff operator. This specific $\overline{g}_{\mu \nu}$-dependence is forced upon us by the requirement of ``background independence''; it has no analog in matter field theories on a non-dynamical spacetime.

We illustrated these issues by means of the CREH truncation which quantizes only one of the degrees of freedom contained in the metric, the conformal factor. If we proceed naively and ignore the special status of the metric we arrive at the $\phi^4$-theory with a negative quartic coupling which, according to Symanzik, is asymptotically free. If, instead, the metric itself is used to set the proper scale of $k$, then the RG flow is different; in particular there exists a non-Gaussian fixed point which is suitable for the asymptotic safety program.

It is quite remarkable that, at the qualitative level, this simple scalar--like theory has exactly the same flow diagram as the full Einstein--Hilbert truncation. It is therefore plausible to conjecture that the complicated selfinteractions of the helicity-2 modes, another feature that distinguishes the metric from matter fields, is possibly not at the heart of asymptotic safety in gravity. Rather, a ``background independent'' quantization scheme seems to be essential.
%
%
%
\pagebreak


\begin{thebibliography}{99}
\bibitem{kiefer}
For a general introduction see C.~Kiefer, \textit{Quantum Gravity}, Second Edition, \\
Oxford Science Publications, Oxford (2007).
%
\bibitem{A}
A.~Ashtekar, \textit{Lectures on non-perturbative canonical gravity},\\
World Scientific, Singapore (1991); \\
A.~Ashtekar and J.~Lewandowski, 
Class.\ Quant.\ Grav.\ 21 (2004) R53.
%
\bibitem{R}
C.~Rovelli, \textit{Quantum Gravity}, Cambridge University Press, Cambridge (2004).
%
\bibitem{T}
Th.~Thiemann, \textit{Modern Canonical Quantum General Relativity},\\
Cambridge University Press, Cambridge (2007).
%
\bibitem{wein}
S.~Weinberg 
in \textit{General Relativity, an Einstein Centenary Survey},\\
S.W.~Hawking and W.~Israel (Eds.), 
Cambridge University Press (1979);\\
S.~Weinberg,
hep-th/9702027.
%
\bibitem{mr}
M.~Reuter, Phys.\ Rev.\ D 57 (1998) 971 and
hep-th/9605030.
%
\bibitem{percadou}
D.~Dou and R.~Percacci, 
Class.\ Quant.\ Grav.\ 15 (1998) 3449.
%
\bibitem{oliver1}
O.~Lauscher and M.~Reuter, 
Phys.\ Rev.\ D 65 (2002) 025013 and hep-th/0108040.
%
\bibitem{frank1}
M.~Reuter and F.~Saueressig, 
Phys.\ Rev.\ D 65 (2002) 065016 and hep-th/0110054.
%
\bibitem{oliver2}
O.~Lauscher and M.~Reuter, 
Phys.\ Rev.\ D 66 (2002) 025026 and hep-th/0205062.
%
\bibitem{oliver3}
O.~Lauscher and M.~Reuter, 
Class.\ Quant.\ Grav.\ 19 (2002) 483 and hep-th/0110021.
%
\bibitem{oliver4}
O.~Lauscher and M.~Reuter, 
Int.\ J.\ Mod.\ Phys.\ A 17 (2002) 993 and hep-th/0112089.
%
\bibitem{souma}
W.~Souma,
Prog.\ Theor.\ Phys.\ 102 (1999) 181.
%
\bibitem{frank2}
M.~Reuter and F.~Saueressig, 
Phys.\ Rev.\ D 66 (2002) 125001 and hep-th/0206145;
Fortschr.\ Phys.\ 52 (2004) 650 and hep-th/0311056.
%
\bibitem{prop}
A.~Bonanno and M.~Reuter,
JHEP 02 (2005) 035 and hep-th/0410191.
%
\bibitem{oliverbook}
For reviews see: M.~Reuter and F.~Saueressig, arXiv:0708.1317 [hep-th],
\\
O.~Lauscher and M.~Reuter in \textit{Quantum Gravity}, B.~Fauser, \\
J.~Tolksdorf and E.~Zeidler (Eds.), Birkh\"auser, Basel (2007) and hep-th/0511260;\\
O.~Lauscher and M.~Reuter in \textit{Approaches to Fundamental Physics}, \\
I.-O.~Stamatescu and E.~Seiler (Eds.), Springer, Berlin (2007).
%
\bibitem{perper1}
R.~Percacci and D.~Perini,
Phys.\ Rev.\ D 67 (2003) 081503;\\
Phys.\ Rev.\ D 68 (2003) 044018;
Class.\ Quant.\ Grav.\ 21 (2004) 5035.
%
\bibitem{codello}
A.~Codello and R.~Percacci,
Phys.\ Rev.\ Lett.\ 97 (2006) 221301;\\
A.~Codello, R.~Percacci and C.~Rahmede,
Int.\ J.\ Mod.\ Phys.\ A 23 (2008) 143;\\
preprint arXiv:0805.2909 [hep-th].
%
\bibitem{litimgrav}
D.~Litim,
Phys.\ Rev.\ Lett.\ 92 (2004) 201301;
AIP Conf.\ Proc.\ 841 (2006) 322;\\
P.~Fischer and D.~Litim, 
Phys.\ Lett.\ B 638 (2006) 497;\\
AIP Conf.\ Proc.\ 861 (2006) 336.
%
\bibitem{frankmach}
P.~Machado and F.~Saueressig,
Phys.\ Rev.\ D 77 (2008) 124045.
%
\bibitem{oliverfrac}
O.~Lauscher and M.~Reuter,
JHEP 10 (2005) 050 and hep-th/0508202.
%
\bibitem{jan1}
M.~Reuter and J.-M.~Schwindt,
JHEP 01 (2006) 070 and hep-th/0511021.
%
\bibitem{jan2}
M.~Reuter and J.-M.~Schwindt,
JHEP 01 (2007) 049 and hep-th/0611294.
%
\bibitem{creh1}
M.~Reuter and H.~Weyer,
preprint arXiv:0801.3287 [hep-th].
%
\bibitem{creh2}
M.~Reuter and H.~Weyer,
preprint arXiv:0804.1475 [hep-th].
%
\bibitem{je1}
J.-E.~Daum and M.~Reuter,
preprint arXiv:0806.3907 [hep-th].
%
\bibitem{max}
P.~Forg\'acs and M.~Niedermaier, 
hep-th/0207028; \\
M.~Niedermaier,
JHEP 12 (2002) 066;
Nucl.\ Phys.\ B 673 (2003) 131;\\
Class.\ Quant.\ Grav.\ 24 (2007) R171.
%
\bibitem{livrev}
For detailed reviews of asymptotic safety in gravity see:\\
M.~Niedermaier and M.~Reuter,
Living Reviews in Relativity 9 (2006) 5;\\
R.~Percacci, arXiv:0709.3851 [hep-th].
%
\bibitem{avact}
C.~Wetterich,
Phys.\ Lett.\ B 301 (1993) 90.
%
\bibitem{ym}
M.~Reuter and C.~Wetterich, \\
Nucl.\ Phys.\ B 417 (1994) 181,
Nucl.\ Phys.\ B 427 (1994) 291, \\
Nucl.\ Phys.\ B 391 (1993) 147, 
Nucl.\ Phys.\ B 408 (1993) 91; \\
M.~Reuter, 
Phys.\ Rev. D 53 (1996) 4430, 
Mod.\ Phys.\ Lett.\ A 12 (1997) 2777.
%
\bibitem{avactrev}
J.~Berges, N.~Tetradis and C.~Wetterich,
Phys.\ Rep.\ 363 (2002) 223;\\
C.~Wetterich,
Int.\ J.\ Mod.\ Phys.\ A 16 (2001) 1951.
%
\bibitem{ymrev}
For reviews of the effective average action in Yang--Mills theory see:\\
M.~Reuter, hep-th/9602012;
J.~Pawlowski, hep-th/0512261;
H.~Gies, hep-ph/0611146.
%
\bibitem{bh}
A.~Bonanno and M.~Reuter, 
Phys.\ Rev.\ D 62 (2000) 043008 and hep-th/0002196;
Phys.\ Rev.\ D 73 (2006) 083005 and hep-th/0602159;\\
Phys.\ Rev.\ D 60 (1999) 084011 and gr-qc/9811026.
%
\bibitem{erick1}
M.~Reuter and E.~Tuiran,
hep-th/0612037.
%
\bibitem{cosmo1}
A.~Bonanno and M.~Reuter,
Phys.\ Rev.\ D 65 (2002) 043508 and hep-th/0106133;
M.~Reuter and F.~Saueressig, 
JCAP 09 (2005) 012 and hep-th/0507167.
%
\bibitem{cosmo2}
A.~Bonanno and M.~Reuter,
Phys.\ Lett.\ B 527 (2002) 9 and astro-ph/0106468; \\
Int.\ J.\ Mod.\ Phys.\ D 13 (2004) 107 and astro-ph/0210472;\\
E.~Bentivegna, A.~Bonanno and M.~Reuter,
JCAP 01 (2004) 001 \\and astro-ph/0303150.
%
\bibitem{entropy}
A.~Bonanno and M.~Reuter,
JCAP 08 (2007) 024 and arXiv:0706.0174 [hep-th].
%
\bibitem{esposito}
A.~Bonanno, G.~Esposito and C.~Rubano,
Gen.\ Rel.\ Grav.\ 35 (2003) 1899;\\
Class.\ Quant.\ Grav.\ 21 (2004) 5005;\\
A.~Bonanno, G.~Esposito, C.~Rubano and P.~Scudellaro,\\
Class.\ Quant.\ Grav.\ 23 (2006) 3103 and 24 (2007) 1443.
%
\bibitem{h1}
M.~Reuter and H.~Weyer, 
Phys.\ Rev.\ D 69 (2004) 104022
and hep-th/0311196.
%
\bibitem{h2}
M.~Reuter and H.~Weyer,
Phys.\ Rev.\ D 70 (2004) 124028
and hep-th/0410117.
%
\bibitem{h3}
M.~Reuter and H.~Weyer,
JCAP 12 (2004) 001 and hep-th/0410119.
%
\bibitem{girelli}
F.~Girelli, S.~Liberati, R.~Percacci and C.~Rahmede,\\
Class.\ Quant.\ Grav.\ 24 (2007) 3995.
%
\bibitem{litim}
D.~Litim and T.~Plehn,
Phys.\ Rev.\ Lett.\ 100 (2008) 131301.
%
\bibitem{mof}
J.~Moffat,
JCAP 05 (2005) 2003; \\
J.R.~Brownstein and J.~Moffat, 
Astrophys.\ J.\ 636 (2006) 721;\\
Mon.\ Not.\ Roy.\ Astron.\ Soc.\ 367 (2006) 527.
%
\bibitem{back}
L.F.~Abbott,
Nucl.\ Phys.\ B 185 (1981) 189;\\
B.S.~DeWitt, 
Phys.\ Rev.\ 162 (1967) 1195;\\
M.T.~Grisaru, P.\ van Nieuwenhuizen and C.C.~Wu,
Phys.\ Rev.\ D 12 (1975) 3203;\\
D.M.~Capper, J.J.~Dulwich and M.\ Ramon Medrano,
Nucl.\ Phys.\ B 254 (1985) 737;\\
S.L.~Adler, 
Rev.\ Mod.\ Phys.\ 54 (1982) 729.
%
\bibitem{floper}
R.~Floreanini and R.~Percacci,
Nucl.\ Phys.\ B 436 (1995) 141;\\
Phys.\ Rev.\ D 46 (1992) 1566.
%
\bibitem{polyakov}
A.M.~Polyakov,
Yad.\ Fiz.\ 64 (2001) 594 \\
$[$English Translation: Phys.\ Atom.\ Nucl.\ 64 (2001) 540$]$.
%
\bibitem{jackiw}
R.~Jackiw, C.~N\'u\~nez and S.-Y.~Pi,
Phys.\ Lett.\ A 347 (2005) 47.
%
\bibitem{syman}
K.~Symanzik,
Nuovo Cim.\ Lett.\ 6 (1973) 77.
%
\bibitem{hist}
For a historic account see: 
G.\ 't Hooft,
Nucl.\ Phys.\ B 254 (1985) 11.
%
\bibitem{kincond}
O.~Lauscher, M.~Reuter and C.~Wetterich,\\
Phys.\ Rev.\ D 62 (2000) 125021 and hep-th/0006099.
%
\bibitem{liouv}
M.~Reuter and C.~Wetterich,
Nucl.\ Phys.\ B 506 (1997) 483 and hep-th/9605039.
%
\bibitem{narpad}
J.V.~Narlikar and T.~Padmanabhan,\\
\textit{Gravity, Gauge Theories and Quantum Cosmology},
D.~Reidel, Dordrecht (1986), Chapter 12 and references therein.
%
\bibitem{opt}
D.~Litim,
Phys.\ Lett.\ B 486 (2000) 92;
Phys.\ Rev.\ D 64 (2001) 105007;\\
Int.\ J.\ Mod.\ Phys.\ A 16 (2001) 2081.
%
\end{thebibliography}
\end{document}